\documentclass[nhess]{copernicus}
\usepackage{amsmath}
\usepackage{lineno} %
\nolinenumbers
\let\linenumbers\nolinenumbers

\AtBeginDocument{\nolinenumbers}

\begin{document}

\title{\vspace*{-3cm} Forecasting rare exceedance events: analytical insights and weather applications}

\Author[a][baggio\_r@univ-corse.fr]{Roberta}{Baggio} %% correspondence author
\Author[a]{Jean-Fran\c{c}ois}{Muzy}
%\Author[]{}{}

\affil[a]{Laboratoire Sciences Pour L'Environnement, UMR 6134, CNRS Université de Corse, Avenue du 9 Septembre, Corte, France}

\runningtitle{Forecasting rare exceedance events}

\runningauthor{R. Baggio, J.F. Muzy}

\received{}
\pubdiscuss{} %% only important for two-stage journals
\revised{}
\accepted{}
\published{}

\firstpage{1}

\maketitle

\begin{abstract}
Accurate short-term forecasting of extreme weather events is important for early warning and risk mitigation. We compare two approaches for predicting site-specific threshold exceedances of weather variables:  direct binary probabilistic models trained on thresholded outcomes and full-distribution parametric models trained on the continuous target. Using an analytically tractable Gaussian random-location model, in which the distribution is predictably shifted by the covariates, we quantify the consequences of the information loss induced by thresholding and derive the rare-event behavior of prediction errors and forecast skill. The analysis predicts an increasing relative advantage of the full-distribution approach as event probability decreases, because binarization progressively discards information contained in the continuous response. We then compare the two approaches
to forecast wind speed and accumulated rainfall at various weather station sites over southeastern France using the same hybrid neural-network architecture. Although wind speed and rainfall depart from the toy-model assumptions, its main qualitative predictions are recovered for both variables: the relative advantage of distributional modeling increases toward rarer thresholds. This agreement further suggests that a substantial fraction of the forecastable signal associated with extreme events arises from predictable shifts in the conditional distribution. For the reasonably suitable parametric families examined, the results show limited sensitivity to the selected class of distribution. Overall, the results highlight the statistical advantage of training on continuous observations rather than on thresholded binary outcomes when forecasting rare threshold exceedances.
\end{abstract}

%\copyrightstatement{TEXT} %% This section is optional and can be used for copyright transfers.

\introduction  %% \introduction[modified heading if necessary]
%%% Context:Importance of extreme event prediction
The accurate and timely prediction of extreme weather events is an important and difficult problem in operational meteorology \citep{IPPC23}. Driven by climate change, the frequency and intensity of localized, high-impact phenomena are clearly increasing, posing severe risks to public safety, civil infrastructure and the stability of renewable energy grids. 

%%% Different Forecasting methods (and their shortcomings?)
Despite remarkable progress in numerical weather prediction (NWP) \citep{Bauer2015}, including convection-permitting systems such as AROME from Météo-France, specifically designed to improve predictions at regional scale \citep{arome2011}, predicting the precise timing and magnitude of localized extremes at the site level remains very challenging. This difficulty mainly stems from the highly non-linear and chaotic nature of atmospheric dynamics \citep{Lorenz1963}, compounded by the smoothing effects of grid-scale parameterizations, unresolved complex topography and sub-grid microphysical processes. 
Alongside  advances in NWP, a long tradition of statistical post-processing has developed to correct systematic forecast biases and quantify predictive uncertainty, initially through Model Output Statistics (MOS) \citep{glahn1972use} and, later, through statistical post-processing of ensemble forecasts \citep{raftery2005using, GneitingEMOS2005}. 
At very short lead times, a complementary tradition of observation-driven nowcasting has emerged to exploit recent measurements and rapidly evolving spatial structures that may not be fully represented by NWP.  Early techniques range from the development of specific stochastic time-series models designed to account for observed localized fluctuations (\cite{BaileMuzyPoggi2011,tascikaraoglu2014review, kaur2023autoregressive}) to optical flow methods for radar tracking \citep{beauchemin1995computation,ayzel2019optical}. While these approaches can provide highly accurate short-term predictions, their skill typically deteriorates with lead time and they may have difficulty representing rapidly evolving weather systems .
%%% ML combines forecasts with obs and offers increased flexibility
Modern machine learning (ML) increasingly bridges these two traditions by enabling nonlinear models that combine recent observations with NWP forecasts and other high-resolution meteorological data. Deep-learning approaches have substantially advanced short-term probabilistic prediction, ranging from radar-based generative nowcasting models such as DGMR \citep{ravuri2021skilful} to architectures such as MetNet and MetNet-2, which integrate heterogeneous meteorological inputs for high-resolution forecasts up to several hours ahead \citep{sonderby2020metnet,espeholt2022deep}. For broader perspectives on machine learning in weather prediction and on the evolution of statistical post-processing toward modern data-driven approaches, see, for instance \citet{Schultz2021} and \citet{vannitsem2021statistical}.
%%% Threshold exceedence approach and proposed methods
This study focuses on the application of ML hybrid approaches to the prediction of threshold exceedances. Here, we use the term hybrid to denote data-driven models that jointly exploit recent local observations and predictors derived from numerical weather prediction forecasts.  In applications such as early warning systems, disaster management or sector-specific planning, high-impact weather conditions are often defined in terms of critical thresholds, such as temperatures above $30^\circ\mathrm{C}$, 1-h precipitation accumulations exceeding $30$ mm or wind speeds above $90$ km/h. Here, we address the site-specific nowcasting of such events at lead times from $1$ to $6$ h, a time range in which rapidly updated probabilistic information is particularly relevant \citep{bojinski2023towards}. More specifically, at a fixed location, for a given atmospheric variable and threshold, our goal is to estimate the corresponding exceedance probability while ensuring that the forecasts can be updated rapidly and remain statistically well calibrated.
From a probabilistic modeling perspective, existing approaches can be broadly divided into two categories. The first is \emph{direct exceedance modeling}, which represents the exceedance or non-exceedance of a given threshold as a Bernoulli outcome and directly estimates its conditional probability $p_t\in[0,1]$. The second category comprises \emph{full-distribution}, or distributional, approaches. Instead of directly estimating the probability of the binary outcome, these methods estimate the conditional distribution of the target variable, often referred to as the predictand. The exceedance probability is then obtained from the corresponding conditional cumulative distribution function (CDF).

%%%%an historical perspectives on these two approaches
Direct probabilistic forecasting of threshold-defined weather events has been widely studied in statistical weather prediction and NWP post-processing. In particular, logistic regression has been extensively used to estimate probabilities of prescribed events from NWP-derived predictors, including in the post-processing of ensemble forecasts and reforecast datasets \citep{hamill2004ensemble,wilks2007comparison,pang2019binary}. In these approaches, the continuous predictand is thresholded to a binary indicator associated with the event of interest. More recently, machine-learning classifiers have generalized this framework by learning flexible nonlinear mappings from heterogeneous predictors to event probabilities, with applications across environmental and meteorological forecasting, including air-pollution exceedances, damaging convective winds, and precipitation exceedances \citep{dutot200724,lagerquist2017machine,schaumann2021calibrated,pujol2025improving}.

The second category of methods, which retains the continuous predictand and models its conditional distribution before thresholding, has also been widely used in forecasting. Full predictive distributions can be represented parametrically, either through a prescribed distribution or through a finite mixture of distributions. Classical approaches include Bayesian Model Averaging and Ensemble Model Output Statistics (EMOS) \citep{raftery2005using,GneitingEMOS2005}, while distributional-regression frameworks allow multiple parameters of a prescribed probability law to depend on atmospheric predictors \citep{Schlosser2019}. These models can be estimated using proper scoring rules such as the logarithmic score or the Continuous Ranked Probability Score (CRPS) \citep{gneiting2007strictly}. More recently, neural distributional regression has extended this framework by using neural networks to learn flexible nonlinear mappings from potentially high-dimensional predictors to distribution parameters, with applications ranging from ensemble post-processing to short-range wind and precipitation forecasting \citep{rasp2018neural,BaggioMuzy2024,baggio2025local,pic2025distributional}.
Other parametric neural density models, such as conditional normalizing flows, learn flexible transformations of a simple reference distribution and thereby avoid prescribing a conventional family such as a Gaussian, Gamma, or Weibull distribution, with recent applications to probabilistic wind-speed forecasting \citep{papamakarios2021normalizing,Marcille24}.
Approaches that do not rely on an explicit parametric density are also available. Nonparametric methods such as Quantile Regression Forests (QRF) estimate conditional distributions without prescribing a parametric family \citep{meinshausen2006quantile,taillardat2016calibrated,park2022learning}, while semiparametric tail models based on Extreme Value Theory (EVT), such as the Peaks Over Threshold approach, model exceedances above a sufficiently high threshold using a Generalized Pareto Distribution \citep{coles2001introduction,friederichs2012forecast}. These approaches involve different assumptions and trade-offs, particularly regarding extrapolation and threshold selection. In the present study, however, we deliberately restrict the full-distribution approach to conditional models based on a conventional parametric family specified in advance.

In light of the preceding discussion, both binary classification and distributional forecasting are well-established approaches to threshold-probability prediction. Comparisons between these strategies have been examined in several studies on the statistical post-processing of weather forecasts where threshold-specific probability models have been contrasted with approaches that estimate full predictive distributions \citep{sloughter2007probabilistic,bentzien2012generating, williams2014comparison,messner2014extending,primo2023comparison}. Related comparisons have also been conducted in other environmental applications \citep{cerqueira2022exceedance}. Previous meteorological studies have reported context-dependent differences in predictive skill, with the contrast between the approaches sometimes becoming more pronounced for rare or extreme events. The contribution of the present study is to isolate and quantify, within an analytically tractable Gaussian location model, the information loss induced by thresholding the continuous predictand. We derive the resulting behavior of exceedance-probability estimation errors and verification scores as event rarity increases and examine the analytical predictions in controlled numerical experiments. The theoretical analysis is complemented by two site-specific applications to strong near-surface winds and intense hourly rainfall in the Mediterranean region of southeastern France. These applications use MeteoNet data for 2016--2018, including co-registered ground-station observations and AROME/ARPEGE outputs over two $550\times550$~km domains encompassing the study area \citep{larvor2021meteonet}. To ensure a direct comparison, both approaches use the same predictors and neural-network backbone in the numerical and meteorological experiments, differing only in their output representation and training objective. Although the meteorological applications do not satisfy all the assumptions of the Gaussian location model and therefore cannot validate its theoretical predictions quantitatively, they provide a useful setting in which to examine whether the qualitative mechanism identified by the model remains visible. Forecast performance is evaluated using the Brier Skill Score and logarithmic score for probabilistic skill, the Area Under the ROC Curve for threshold-independent discrimination, and the Peirce Skill Score for binary discrimination. The sensitivity of the distributional approach to the choice of parametric family is also investigated.

The paper is organized as follows. Section~\ref{sec:prob_statement} formulates the forecasting problem, introduces the direct-classification and full-distribution approaches and defines the verification metrics. Section~\ref{sec:theormod} develops the theoretical framework and presents numerical experiments examining the rare-event behavior of both approaches within a simplified generative model. Section~\ref{sec:application} applies both approaches to MeteoNet data and presents the neural-network architecture and empirical results for wind speed and hourly accumulated rainfall. Section~\ref{sec:conclusion} summarizes the main findings and outlines directions for future research. The appendices provide the technical derivations and additional methodological details.

\section{Statement of the problem}\label{sec:prob_statement}

\subsection{Extreme events as binary events}\label{sec:extremebins}
\label{sec:prediction}
In this section, we introduce the main notations used throughout the paper and formally define the forecasting problem. Let $Y_t$ denote the value at time $t$ of an atmospheric variable, such as wind speed, temperature or hourly accumulated rainfall, at a given location. We model $\{Y_t\}$ as a stationary stochastic process taking values in $\mathbb R$.

At any time $t$, given a threshold $Y_*$ and a time horizon $h>0$, our objective is to predict exceedance events of $Y_{t+h}$. We formalize such events using a binary indicator:
\begin{equation}
    \label{eq:def_Iq}
    I_{t+h}(Y_*) = \mathcal{H}\left(Y_{t+h} - Y_*\right) =
    \begin{cases}
        1 & \text{if } Y_{t+h} \geq Y_*, \\
        0 & \text{otherwise},
    \end{cases}
\end{equation}
where $\mathcal{H}$ denotes the Heaviside step function. Predicting extreme events thus reduces to forecasting $I_{t+h}(Y_*)$ for large $Y_*$ values, particularly those corresponding to high percentiles of the site's climatological distribution. For a probability level $1-p$, we define the associated quantile $Q_p$ as:
\begin{equation}
    \label{eq:def_quantile}
    F_C(Q_p) = 1 - p,
\end{equation}
where $F_C(z) = \int_{-\infty}^z f_C(u) \, du$ is the cumulative distribution function (CDF) derived from the climatological probability density function $f_C(u)$. When $p \ll 1$, $I_{t+h}(Q_p)$ indicates whether $Y_{t+h}$ exceeds a threshold chosen in the distribution's upper tail.  In the remainder of this paper, we drop the explicit threshold dependency and implicitly let $I_{t+h}$ denote $I_{t+h}(Q_p)$ unless stated otherwise. Naturally the same mathematical framework can be applied to the complementary problem of non-exceedance below a threshold in the lower tail. In the present work, however, we restrict attention to extreme weather events associated with unusually high values of the variable of interest and therefore focus exclusively on upper-tail exceedances.
The previous prediction task can be formulated as a probabilistic binary classification problem. Given an observable covariate vector $X_t$ representing the information available at time $t$, the quantity to be estimated is the conditional exceedance probability 
\begin{equation} 
\label{eq:defp} 
p_t \underset{\mathrm{def}}{=} \Pr\left(I_{t+h}=1\mid X_t\right). 
\end{equation}
A \emph{probabilistic prediction}, namely an estimate of $p_t$, denoted as ${\widehat{p}}_t$, can then be converted to a \emph{deterministic prediction} by thresholding at $p^\star$:
\begin{equation}
\label{eq:det_pred}
\widehat{I}_{t+h} =
\begin{cases}
1 & \text{if } {\widehat{p}}_t > p^\star \; , \\
0 & \text{otherwise}.
\end{cases}
\end{equation}

In the following sections, we introduce two distinct methodologies to estimate the conditional probability $p_{t}$ and review the probabilistic and deterministic verification metrics used to assess and compare the quality of these threshold exceedance forecasts.

\subsection{Modeling approaches}
Our objective is to compare two distinct model classes for estimating $p_t$ defined in Eq.~(\ref{eq:defp}). Both approaches utilize observable covariates $X_t$ as input features representing the information available at time $t$. However, they fundamentally differ in how they process the target variable during training.

\paragraph{Class $\mathcal{M}_1$: Direct probability estimation}
The first approach ($\mathcal{M}_1$) treats the task strictly as a binary classification problem. The class $\mathcal{M}_1$ is the set of all models $M_1$ that directly map the covariates to the estimated exceedance probability:
\begin{equation}
    \label{eq:direct_estimation}
    M_1(X_t; \widehat{{\boldsymbol{\theta}}}) = \widehat{p}_t^{(1)},
\end{equation}
where $\widehat{{\boldsymbol{\theta}}}$ represents the learned model parameters.  
This formulation corresponds to standard probabilistic classification, where parameters are typically, but not exclusively, optimized using the Binary Cross-Entropy (BCE) loss:
\begin{equation}
    \label{eq:bce_loss}
    \mathcal{L}_{\text{BCE}} =  -\frac{1}{N} \sum_{i=1}^N \left[ I_{t_i+h} \ln\left(\widehat{p}_{t_i}^{(1)}\right) + \left(1 - I_{t_i+h}\right) \ln\left(1 - \widehat{p}_{t_i}^{(1)}\right) \right],
\end{equation}
where $\{t_i\}_{i=1\ldots N}$ identify the times of the sample associated with training set.  The BCE is the negative average Bernoulli log-likelihood, equivalently the average logarithmic score for binary events, which is a strictly proper scoring rule \citep{gneiting2007strictly}. 
Regardless of the particular classification model, the training procedure involves only the binary indicators $I_{t_i+h}$. Consequently, the continuous magnitude information contained in $Y_{t_i+h}$ is discarded and only whether the threshold was exceeded is retained. As mentioned in the Introduction, logistic regression provides a classical example of this approach whereas neural-network classifiers allow for more flexible nonlinear implementations. 

\paragraph{Class $\mathcal{M}_2$: Distribution-based probability estimation} 
The second class $\mathcal{M}_2$ consists of models $M_2$ that produce exceedance forecasts through a two-stage procedure: (i) they estimate a parametric conditional distribution of $Y_{t+h}$ given $X_t$, and (ii) they derive the exceedance probability from the corresponding conditional CDF. 
Let $f(y;\Pi)$ denote the density of the chosen parametric family, used to model the conditional distribution of $Y_{t+h}$ given $X_t$. At this stage, this family is not assumed to contain the true conditional distribution. 
The data are used to estimate the time-dependent parameter vector $\Pi_t$ through 
\begin{equation} 
\label{eq:param_estimation} 
M_2\left(X_t;\widehat{\boldsymbol{\theta}}\right) = \widehat{\Pi}_t. 
\end{equation} 
For example, in the Gaussian location model considered in Section~\ref{sec:theormod}, 
the conditional variance is known and constant, while $\Pi_t=\mu_t$. Consequently  
$M_2\left(X_t;\widehat{\boldsymbol{\theta}}\right) = \widehat\mu_t$, where $\widehat\mu_t$ denotes the estimated conditional mean. 

A common method for estimating $\widehat{\boldsymbol{\theta}}$ consists of minimizing the negative log-likelihood of the continuous observations: 
\begin{equation} 
\label{eq:log_likelihood} 
\mathcal L_{\mathrm{LL}} = -\frac{1}{N} \sum_{i=1}^{N} \ln f\left( Y_{t_i+h}; \widehat{\Pi}_{t_i} \right). 
\end{equation} 
In contrast to $\mathcal M_1$, the training loss for $\mathcal M_2$ uses the continuous values of $Y_{t+h}$ and therefore all training observations, independently of the selected exceedance threshold. Once $\widehat{\Pi}_t$ has been estimated, the exceedance probability is computed analytically or numerically as 
\begin{equation} 
\label{eq:prob_from_dist} 
{\widehat p}_t^{(2)} = \int_{Q_p}^{\infty} f\left(y;\widehat{\Pi}_t\right)\,dy. 
\end{equation} 
Thus, a single fitted conditional distribution provides exceedance probabilities for arbitrary thresholds.

\subsection{Forecast verification of binary events}
Evaluating threshold-exceedance forecasts requires both deterministic metrics to assess binary decision-making and probabilistic scores to evaluate the quality of the predicted probabilities, including their calibration and resolution. Since we focus on rare extreme events, the chosen metrics should remain informative under very low event base rates.  In other words, the event and non-event classes are characterized by severe class imbalance \citep{Jolliffe2004,wilks2011statistical}. In the following, we introduce the deterministic and probabilistic verification metrics used throughout this study. Taken together, the selected metrics provide a base-rate-insensitive measure of discrimination, which is particularly important for rare-event forecasting and a principled evaluation of probabilistic forecast quality based on proper scoring rules.

\paragraph{Deterministic predictions}
Binary forecasts $\widehat{I}_t(Y_*)$ are evaluated using standard metrics derived from the contingency table which summarizes the four possible forecast-observation outcomes: true positives (TP; event forecast and observed), true negatives (TN; non-event forecast and observed), false positives (FP; event forecast but not observed), and false negatives (FN; non-event forecast but event observed).

As primary measure of binary discrimination, we consider the Peirce Skill Score (PSS, \citet{wilks2011statistical}):
\begin{equation}
\text{PSS} = \frac{\mathrm{TP}}{\mathrm{TP}+\mathrm{FN}} - \frac{\mathrm{FP}}{\mathrm{TN}+\mathrm{FP}}.
\label{eq:PSS}
\end{equation}
PSS ranges from $−1$ to $1$, with $\text{PSS} = 1$ corresponding to a perfect forecast and $\text{PSS} = 0$ corresponding to unskilled forecasts like a purely random or constant predictions. PSS is an equitable score in the sense that it rates equally to 0 uninformative forecasts \citep{wilks2011statistical} and has no explicit dependence on the event base rate.  From a societal perspective, PSS is suitable for comparing discrimination of forecasting systems \citep{ebert2021methodological} in sense that it directly quantifies their ability to detect potentially harmful extreme events while limiting false alarms, two essential requirements for reliable early-warning and civil-protection decisions. Note that, for a calibrated probability forecast the  threshold in  Eq.~\eqref{eq:det_pred}  that maximizes the PSS is simply the climatological event probability $p$ \citep{mason1979reducing, Jolliffe2004}.
For completeness, we also evaluate the forecasts of two alternative popular scores, the Heidke Skill Score (HSS) and Critical Success Index (CSI). Unlike the PSS, these scores depend explicitly on event prevalence or forecast marginal frequencies, making their comparison across increasingly rare thresholds less directly interpretable. Their definitions and corresponding results are therefore reported in Appendix \ref{app:csi_hss} as complementary diagnostics.

\paragraph{Probabilistic predictions}
We evaluate the probabilistic forecasts using  the Brier Score (BS) and the logarithmic score (LS), which are  \emph{strictly proper} scoring rules  \citep{gneiting2007strictly}. By definition, the expected value of a strictly proper scoring rule is uniquely optimized when the forecast probability equals the true conditional probability. BS measures the mean squared error of the forecast probabilities:
\begin{equation}
\label{eq:def_Brier}
\text{BS} = \frac{1}{N} \sum_{i=1}^N \left( \widehat{p}_{t_i} - I_{t_i+h} \right)^2.
\end{equation}
The BS is negatively oriented, with smaller values indicating better probabilistic forecasts. To express performance relative to climatology, whose BS is  $p(1-p)$, we use the Brier Skill Score (BSS):
\begin{equation}
\label{eq:def_BSS}
\text{BSS} = 1 - \frac{\text{BS}}{p(1-p)} = \frac{\mathrm{Res} - \mathrm{Rel}}{U},
\end{equation}
where $\mathrm{Rel}$ and $\mathrm{Res}$ denote the reliability and resolution components, and $U=p(1-p)$ represents the climatological uncertainty \citep[see][for full algebraic decompositions]{murphy1973, Jolliffe2004, wilks2011statistical}. The BSS is positively oriented, with a maximum value of 1, while negative values indicate performance worse than climatology. LS simply corresponds to $\mathcal L_{\mathrm{BCE}}$ in Eq.~\eqref{eq:bce_loss}, which is commonly used to train models in class $\mathcal M_1$: 
\begin{equation} 
\label{eq:def_logScore} 
\mathrm{LS} = -\frac{1}{N} \sum_{i=1}^{N} \left[ I_{t_i+h}\ln\widehat p_{t_i} + (1-I_{t_i+h})\ln(1-\widehat p_{t_i}) \right]. 
\end{equation} 
Like the BS, the LS is negatively oriented: smaller values indicate better forecasts, with zero corresponding to a perfect forecast. Although a climatology-referenced skill score could also be defined for the LS, we use the additive difference between the two models because the expected LS decomposes into the entropy of the true Bernoulli distribution and a Kullback--Leibler divergence. Consequently, $\Delta \mathrm{LS}_{12} = \mathrm{LS}_1-\mathrm{LS}_2$ directly compares the corresponding excess risks (see Appendix \ref{app:logscore}). The evaluation of probabilistic forecasts can be complemented by the Area Under the ROC Curve (AUC), which summarizes discrimination over all possible decision thresholds \citep{Jolliffe2004}. The AUC ranges from 0 to 1, with 0.5 corresponding to no discrimination and 1 to perfect discrimination, and has no explicit dependence on the event base rate.

\section{Theoretical analysis and numerical experiments with a Gaussian location model}\label{sec:theormod}

To compare the two modeling strategies within a theoretical framework that permits explicit calculations, we consider a Gaussian location model inspired by \citet{lerch2017forecaster}. In this setting, $Y_{t+h}\mid X_t$ follows a Gaussian distribution with a covariate-dependent mean $\mu(X_t)$ and a known, constant conditional variance. The full-distribution model is assumed to be correctly specified, allowing us to isolate the information loss caused by replacing the continuous response with its thresholded binary indicator. The analysis is conducted within a regular large-sample framework, assuming consistency, asymptotic normality and negligible squared bias. General expressions for the errors associated with both approaches are obtained using the delta method, while closed-form approximations require additional simplifying assumptions detailed in Appendix~\ref{app:asympt_error}. Although this framework does not capture the full complexity of the model classes $\mathcal M_1$ and $\mathcal M_2$, it provides a controlled setting in which the statistical consequences of response thresholding can be quantified explicitly. A central objective is to formalize and quantify the intuition that, as events become rarer, the binary response carries progressively less information for estimating the exceedance probability, whereas $M_2$ continues to exploit information contained in all the observed values $Y_{t+h}$. We therefore derive the rare-event behavior of probability-estimation errors and forecast verification scores as the climatological exceedance probability $p$ approaches zero. These asymptotic results quantify how the difference in available information affects the relative performance of the two approaches. Finally, a numerical experiment based on a specific choice of $\mu(X_t)$ is used to assess whether the predicted behaviors are recovered when $M_1$ and $M_2$ are implemented using simple neural networks.

\subsection{Estimation of the asymptotic errors for each model and their effects on performance scores.}
\label{ss:analytical}
Let $(X_t)_{t}$ be a $d$-dimensional stationary random process of observable covariates. An unknown mapping $F: \mathbb{R}^d \to \mathbb{R}$ generates a latent mean signal:
\begin{equation}
\label{eq:def_mu}
\mu_t = F(X_t) \; .
\end{equation}
By stationarity of $X_t$, the process $ \mu_t $ is also stationary and we assume its marginal distribution is Gaussian with mean zero and variance $s^2$. At time $t+h$, we observe:
\begin{equation}
    Y_{t+h} = \mu_t + \nu_{t+h}, 
    %\quad \nu_t \overset{\text{i.i.d.}}{\sim} \phi_{\sigma^2},
    \label{eq:observation_model}
\end{equation}
where $\nu_t$ is a white noise process of variance $\sigma^2$ that is also assumed to be Gaussian. In that respect, the law of $Y_t$ is Gaussian of variance $\sigma^2_Y = s^2+\sigma^2$.
Throughout this paper, we will denote by $\phi(z)$ the standard normal density and $\Phi(z)$ the associated cumulative distribution function (CDF) and $\Phi^{-1}(q)$ the inverse cumulative distribution, namely, the inverse function of $\Phi(z)$. If $Z = (Z_1,\ldots,Z_n)$ is a random vector of law $f_{Z}(z_1,\ldots,z_n)$, the expectation of any function 
$G(z) = G(z_1,\ldots,z_n)$ with respect to $Z$, is denoted as 
\begin{equation}
     %\mathbb{E}_{Z} \left[G(z) \right] =
     \mathbb{E}_{Z} \left[G(Z) \right] =
\int dz_1 \ldots dz_n \; G(z_1,\ldots,z_n) \;  f_{Z}(z_1,\ldots,z_n) \; .
\end{equation}
For a fixed threshold $Y_* \in \mathbb{R}$, the quantity of interest is the conditional exceedance probability:
\begin{equation}
     P(X_t)  \underset{\mathrm{def}}{=} p_t =
    \Phi \left(\frac{\mu_t-Q_p}{\sigma}\right) \; .
    \label{eq:true_pt}
\end{equation}
where we noticed that $\mathrm{Pr} \left(\nu >z \right) = 1-\mathrm{Pr} \left(\nu\leq z \right) = 1 - \Phi(\frac{z}{\sigma})= \Phi(-\frac{z}{\sigma})$. We focus on the small-probability regime, namely $p \ll 1$ and we notice that for our model, the quantile $Q_p$ reads:
\begin{equation}
    Q_p = -\sqrt{s^2 + \sigma^2} \cdot \Phi^{-1}(p).
    \label{eq:def_X0}
\end{equation}
A key parameter of the model is the noise-to-signal ratio:
\begin{equation}
	\label{eq:ns_ratio}
	\eta^2 = \frac{\sigma^2 }{s^2},
\end{equation}
representing the relative magnitude of the idiosyncratic noise fluctuations $\nu_{t+h}$ in $Y_{t+h}$ compared to its conditional mean signal $\mu_t$. When $\eta^2$ is small, the predictability of $I_{t+h}$ is high because
$p_t$ varies between values close to either $p_t = 0$ or $p_t = 1$ whereas when $\eta^2 \to \infty$, the predictability is small since $p_t$ variation around its mean value $p$ are small. 
In this context, it is natural to model $\eta^2$ as an increasing function of the forecasting horizon $h$, reflecting the fact that predictability inherently decreases over longer horizons. Consequently, when comparing model output to empirical results, a larger horizon must correspond to a higher effective value of $\eta^2$. Let us remark that, with little algebra, one can easily compute the first two moments of $p_t$:
\begin{equation}
\label{E1}
    \mathbb{E}_{X_t}[p_t] = \mathbb{E}_{X_t}\left[ \Phi\left( \frac{\mu_t - Q_p}{\sigma} \right) \right]  = p.
\end{equation}
and
\begin{equation}
\label{E2}
 \mathbb{E}_{X_t}[p_t^2] = \mathbb{E}_{X_t}\left[ \Phi\left( \frac{\mu_t - Q_p}{\sigma} \right)^2 \right] = \Phi_2\left( \Phi^{-1}(p) ; r \right)
\end{equation}
where $ r = \frac{1}{1+\eta^2} $ and  $\Phi_2(x ; r)$ 
stands for the cumulative distribution function of the standard bivariate normal distribution with correlation $r$ (see definition (\ref{eq:Phi_k_diagonal})).
Notice that, when $\eta^2 \to \infty$, $r \to 0$.  Since $\Phi_2(q,0) = \Phi^2(q)$, we thus have, when $\eta^2 \to \infty$, $\mathrm{Var}(p_t) = {\mathbb{E}}(p_t^2) - {\mathbb{E}}(p_t)^2 \to p^2 - p^2 = 0$. Indeed, as mentioned above, when $\eta \to \infty$, $Y_t$, is a pure, unpredictable, Gaussian white noise of variance $\sigma^2$ and therefore $p_t$ is constant, $p_t = p$ independently of $t$. On the other hand, when $\eta^2 \to 0$, $r \to 1$ and since $\Phi_2(q,1) = \Phi(q)$,
one has $\mathrm{Var}(p_t) = {\mathbb{E}}(p_t^2) - {\mathbb{E}}(p_t)^2 \to p - p^2 = p(1-p)$ which is the variance of a Bernoulli process. This is simple to understand since, in that case, $Y_{t+h}$ reduces to its predictable component $\mu_{t}$ and $p_t$ becomes itself a Bernoulli process since $p_t = 1$ with probability $p$ (if $\mu_{t} \geq Q_p$) and $p_t = 0$ otherwise.

Our goal is to estimate the function $P(\cdot)$ in Eq. ~(\ref{eq:true_pt}) which maps $X_t$ to the conditional probability $p_t$. This can be done using a model  $M(.;{\boldsymbol{\theta}})$, whose parameters are learned over a training set $\{(X_{t_i}, Y_{t_i+h})\}_{i=1}^N$. We first consider a model $M_1(.;{\boldsymbol{\theta}})$ that, following  $\mathcal{M}_1$ approach, directly outputs an estimate $\widehat{p}_t^{(1)}$ of $p_t$. Its best parameters are obtained by minimizing the binary cross-entropy (BCE) loss associated with observed exceedances $I_{t+h}(Q_p)$. We also consider a model $M_2(.;{\boldsymbol{\theta}})$ in the class $\mathcal{M}_2$ which provides an estimation of $\mu_t$, denoted as ${\hat \mu_t}$, allowing
one to compute the conditional probability estimate $\widehat{p}_t^{(2)}$ using Eq. ~(\ref{eq:true_pt}). $M_2$ model's parameters are obtained by  maximizing the log-likelihood which reduces, as $\sigma^2$ is known, to the Mean Squared Error.  

In Appendix~\ref{app:asympt_error}, we derive the prediction errors under standard large-sample regularity conditions. More precisely, we assume that the estimators are consistent and asymptotically normal, with asymptotic covariance given by the inverse Fisher information matrix. The full Fisher matrices are retained, including their off-diagonal entries.
To obtain tractable expressions, the local variance functions are approximated by positive constants.
More general covariate-dependent forms can be treated analytically but are not required to reproduce the main numerical trends reported in the next section. 
For $M_1$, we additionally factorize the Fisher information by neglecting the matrix-valued covariance between the conditional Bernoulli variance factor $p_t(1-p_t)$ and the local sensitivity matrix. 
This approximation retains the full parameter covariance structure and gives 
\begin{equation} 
\label{eq:error_M1_finite_main} 
{\cal E}_1 \simeq \frac{V_1}{N} \frac{\Phi_2(\Phi^{-1}(p);r)} {p-\Phi_2(\Phi^{-1}(p);r)}, 
\end{equation} 
where $\Phi_2(z,r)$ is defined in Eq.~(\ref{eq:Phi_k_diagonal}), $r=\frac{1}{1+\eta^2}$ and $\simeq$ 
indicates that the expression is obtained under the simplifying approximations stated above (see Appendix~\ref{app:asympt_error}).
For $M_2$, the corresponding constant-local-variance approximation is 
\begin{equation} \label{eq:error_M2_finite_main} {\cal E}_2 \simeq \frac{V_2\eta} {2\pi N\sqrt{2+\eta^2}} \exp\left[ -\frac{1+\eta^2}{2+\eta^2} \left\{\Phi^{-1}(p)\right\}^2 \right]. 
\end{equation} 
These large-sample expressions are first obtained for fixed $p>0$ as $N\to\infty$ and their rare-event behavior is then examined as $p\to0$ or, more generally, if $p=p_N$ decreases jointly with $N$, as $Np_N\to\infty$ (see Appendix~\ref{app:asympt_error}).
In the rare-event regime, the Gaussian tail equivalents derived in
Appendix~\ref{app:asympt_error} yield
\begin{align}
\label{eq:L2_error_M1}
{\cal E}_1
&\sim
\frac{K_1(\eta)}{N}
p^{\frac{\eta^2}{2+\eta^2}}
\left[\ln(1/p)\right]^{
-\frac{1}{2+\eta^2}},
\\
\label{eq:L2_error_M2}
{\cal E}_2
&\sim
\frac{K_2(\eta)}{N}
p^{\frac{2+2\eta^2}{2+\eta^2}}
\left[\ln(1/p)\right]^{
\frac{1+\eta^2}{2+\eta^2}},
\end{align}
where $f(p)\sim g(p)$ means that $f(p)/g(p)\to1$ as $p\to0$. The positive prefactors $K_1(\eta)$ and $K_2(\eta)$ 
are independent of $p$ for fixed $\eta$, but contain the model-dependent local variance factors $V_1$ and $V_2$, respectively. 
The expectations defining $\mathcal E_1$ and $\mathcal E_2$ are taken jointly over repeated training samples and the covariate distribution. 
Under the negligible-bias assumption, these errors correspond to the integrated asymptotic variances of the probability estimators.
At fixed, sufficiently small $p$, the finite-$p$ expressions derived in Appendix~\ref{app:asympt_error} show that both errors decrease in the large-noise regime and approach finite limits as $\eta\to\infty$. This behavior is also observed in the numerical experiments below. This apparently counterintuitive behavior can be explained by the decreasing intrinsic variability of the conditional probability $p_t$, since $\operatorname{Var}(p_t) = \mathbb E[p_t^2]-p^2 = \Phi_2(z_p;r)-p^2$ with $r=(1+\eta^2)^{-1}$. As $\eta^2$ increases, the latent correlation $r$ tends to zero, reflecting the increasing dominance of the idiosyncratic noise over the predictable component $\mu_t$. Consequently, $\Phi_2(z_p;r)$ tends to the product of the marginal probabilities $p^2$, and $\operatorname{Var}(p_t)$ tends to zero. The conditional probability $p_t$ therefore approaches the constant marginal probability $p$, making $p_t$ easier to estimate in absolute terms, although progressively less informative for discriminating events from non-events.

Equations~(\ref{eq:L2_error_M1}) and (\ref{eq:L2_error_M2}) also give, at fixed $\eta$, as $p \to 0$,
\begin{equation} 
\label{eq:error_ratio_main} 
\frac{{\cal E}_2}{{\cal E}_1} \sim \frac{K_2(\eta)}{K_1(\eta)} p\ln(1/p). 
\end{equation} 
Thus, at the level of the leading algebraic powers of $p$, the distributional approach gains one power of $p$ over direct classification. This difference reflects the loss of Fisher information caused by reducing the continuous observations to a rare Bernoulli event: $M_2$ estimates the conditional-distribution parameter from all $N$ continuous values, whereas the scalar Fisher information available to $M_1$ is proportional to $N[p-\Phi_2(z_p;r)] \simeq Np$. These results are specific to the regular Gaussian location model with
known conditional variance, correct specification, asymptotic
normality and negligible squared bias. More general local variance
functions can be treated analytically, whereas model misspecification,
non-zero bias or failure of the Fisher-information factorization
for $M_1$ requires numerical evaluation of the general expressions in
Appendix~\ref{app:asympt_error}.

The previous considerations are complemented by evaluating forecast
performance using verification scores, namely the PSS for binary
predictions, the BSS and LS for probabilistic predictions.
The use of the BSS rather than the raw BS is particularly useful in the present context, 
since normalization with respect to climatology removes the leading prevalence-driven scaling of the BS and 
makes the evolution of probabilistic skill across increasingly rare events more interpretable. 
Appendices~\ref{app:logscore} and \ref{app:pss} relate the probability-estimation errors $\mathcal E_1$ and $\mathcal E_2$ to the BSS, LS and to the PSS, respectively. In particular, Eq.~\eqref{BSS_k_vs_E_k} expresses the BSS directly in terms of the probability-estimation error $\mathcal E_k$, for $k=1,2$. Using the
Gaussian tail behavior derived in Appendix~\ref{app:logscore}, we obtain
\begin{equation}
\label{eq:BSS_k_asympt}
\mathrm{BSS}_k
\sim
C(r)
p^{\frac{\eta^2}{2+\eta^2}}
\left[\ln(1/p)\right]^{-\frac{1}{2+\eta^2}}
-
\frac{\mathcal E_k}{p}.
\end{equation}
Since $\mathcal E_2/\mathcal E_1\to0$ as $p\to0$ according to
Eq.~\eqref{eq:error_ratio_main}, the estimation-error penalty is
asymptotically smaller for $M_2$, leading to
$\mathrm{BSS}_2>\mathrm{BSS}_1$ for sufficiently rare events.

We can also compare the two methods in terms of LS. An additive comparison is particularly natural for the logarithmic score, since its expected value can be decomposed into the entropy of the true distribution and a Kullback-Leibler divergence. Consequently, $\Delta \mathrm{LS}_{1,2}= \mathrm{LS}_1 - \mathrm{LS}_2$  directly corresponds to the difference between the KL divergences associated with the two models \citep{weijs2010kullback}.
We demonstrate in Appendix \ref{app:logscore} that it reads, when $p \downarrow 0$:
\begin{equation}
\label{eq:delta_LS_main}
\Delta\mathrm{LS}_{1,2}
\simeq
\frac{1}{2N}
\left[
V_1
-
\frac{2V_2\eta^2}{1+\eta^2}
p\ln\left(\frac{1}{p}\right)
\right].
\end{equation}
Hence, for sufficiently small $p$,
$\Delta\mathrm{LS}_{1,2}>0$, indicating that $M_2$ outperforms $M_1$
according to the logarithmic score.

For the PSS, Eqs.~\eqref{eq:pss1_a} and \eqref{eq:pss2_a} in
Appendix~\ref{app:pss} express the scores as expectations over the
distribution of $p_t$, which can be evaluated numerically. In the
regular joint regime $N\to\infty$, $p\to0$ and $Np\to\infty$, their
asymptotic behaviors, established in
Eqs.~\eqref{eq:pss_1_asympt_appendix} and
\eqref{eq:pss_2_asympt_appendix}, are
\begin{align}
\label{eq:pss_1_th}
\mathrm{PSS}_1
&\simeq
1-K_\eta p^{\kappa^2}
-\frac{D_{1,\eta}}{N}p^{-\gamma},
\\
\label{eq:pss_2_th}
\mathrm{PSS}_2
&\simeq
1-K_\eta p^{\kappa^2}
-\frac{D_{2,\eta}}{N}p^{1-\gamma},
\end{align}
where $K_\eta,D_{1,\eta},D_{2,\eta}>0$ are independent of $p$ for
fixed $\eta$,
\begin{equation}
\gamma
=
\frac{2\eta}{\sqrt{1+\eta^2}+\eta}
\in(0,1),
\qquad
\kappa
=
\sqrt{1+\eta^2}-\eta,
\end{equation}
and $D_{1,\eta}$ and $D_{2,\eta}$ contain the model-dependent local
variance factors $V_1$ and $V_2$, respectively. The common term $1-K_\eta p^{\kappa^2}$ represents the asymptotic PSS based on the true conditional probability $p_t$. For fixed finite $\eta$, it tends to one as $p\to0$. At fixed $p$, however, intrinsic discrimination decreases as $\eta$ increases and vanishes in the pure-noise limit. Equation~\eqref{eq:pss_2_th} predicts that $\mathrm{PSS}_2$ increases
as $p$ decreases. Indeed, high thresholds are exceeded predominantly
when the predictable component $\mu_t$ is large, reducing the relative
influence of the idiosyncratic noise $\nu_{t+h}$ in
Eq.~\eqref{eq:observation_model}. For $M_1$, this gain in intrinsic
predictability is offset by the loss of information associated with
increasingly rare binary labels, as reflected by the correction
$N^{-1}p^{-\gamma}$ in Eq.~\eqref{eq:pss_1_th}. For sufficiently small $p$, the performance ratio satisfies
\begin{equation}
\begin{aligned}
\label{eq:PSS_ratio_main}
\frac{\mathrm{PSS}_1}{\mathrm{PSS}_2} \; \; \;
& \simeq &
1-\frac{A_{p,\eta}}{N}p^{-\gamma}
+
\mathcal O\left(
\frac{p^{-2\gamma}}{N^2}
\right), \; \mathrm{with} \\
A_{p,\eta} \; \;
& = &
D_{1,\eta}
\left[
1+K_\eta p^{\kappa^2}
+o\left(p^{\kappa^2}\right)
\right]
>0,
\end{aligned}
\end{equation}
where $0<\kappa^2<1$ was used. Hence,
$\mathrm{PSS}_1/\mathrm{PSS}_2<1$, showing that $M_2$ outperforms
$M_1$, with a relative PSS difference of order
$(Np^\gamma)^{-1}$. Since $0<\gamma<1$, $Np\to\infty$ also implies
$Np^\gamma\to\infty$, so that this difference remains asymptotically
small. At fixed $p$, both methods converge as $N\to\infty$ to the same
PSS based on the true conditional probability $p_t$.

\subsection{Numerical validation using a toy generative model}
To empirically validate our analytical findings and provide illustrative examples, we implement the simple model described in Appendix~\ref{app:model1}. Specifically, according to Eq. ~(\ref{eq:def_harm}), the latent process $\mu_t $ (Eq.~(\ref{eq:def_mu})) is constructed as a weighted sum of harmonic functions applied to the components of a $d$-dimensional Gaussian white noise input $X_t$. 
The resulting process $\mu_t$ is zero mean, approximately normal and the weights chosen such that its variance is $s^2$.

Both estimation models $M_1(X_t; {{\boldsymbol{\theta}}})$ and $M_2(X_t; {{\boldsymbol{\theta}}})$ employ identical multi-layer perceptron (MLP) architectures, each with 3 layers featuring:
\begin{itemize}
	  \item[-] Input dimension matching the $d$-dimensional covariates $X_t$
    \item[-] Two hidden layers with 32 ReLU-activated units
    \item[-] Linear output layers representing logits for $M_1$ and regression for $M_2$.
\end{itemize}

The $\mathcal{M}_1$-type model $M_1$ directly estimates exceedance probabilities $\mathrm{Pr}(Y_{t+h} > Y_* \mid X_t)$ using sigmoid output activation and minimizing the binary cross-entropy loss to optimize ${\boldsymbol{\theta}}$, while the $\mathcal{M}_2$-type model $M_2$ predicts the latent process $\mu_t$ by minimizing the mean squared error (MSE) between predicted and observed $Y_{t+h}$ values. 
Both models are trained using the Adam optimizer with a batch size of $2^{12} = 4096$, learning rate of $10^{-3}$ and early stopping based on validation loss with a patience of 20 epochs. The validation set comprises a separate 10\% split of the original training data.

Our experimental setup uses $d = 12$ input dimensions with training and test sets containing $N = 2^{16} $ and $N' = 2^{14}$  samples respectively. 
The value $d=12$ was chosen as a compromise between obtaining an accurate Gaussian approximation for the latent signal and limiting the number of model parameters and the computational cost. Additional experiments with $d=10$ and $d=18$ produced qualitatively similar results, 
indicating that the main conclusions are not sensitive to this choice. All simulations, model training and predictions were implemented in Python using the PyTorch framework, ensuring efficient GPU acceleration and reproducible results. To robustly evaluate estimation errors, we use a Monte Carlo resampling procedure: While keeping the set of test pairs $\{(Y_{t_i+h}, X_{t_i})\}_{i=1; \ldots, N'}$ constant, we train both models on 30 independent realizations of the training set. This methodology provides stable estimates of model performance (notably their bias and variance) while accounting for the inherent variability in training process.

\begin{figure*}[h]
	\centering
	\includegraphics[width=0.9\linewidth]{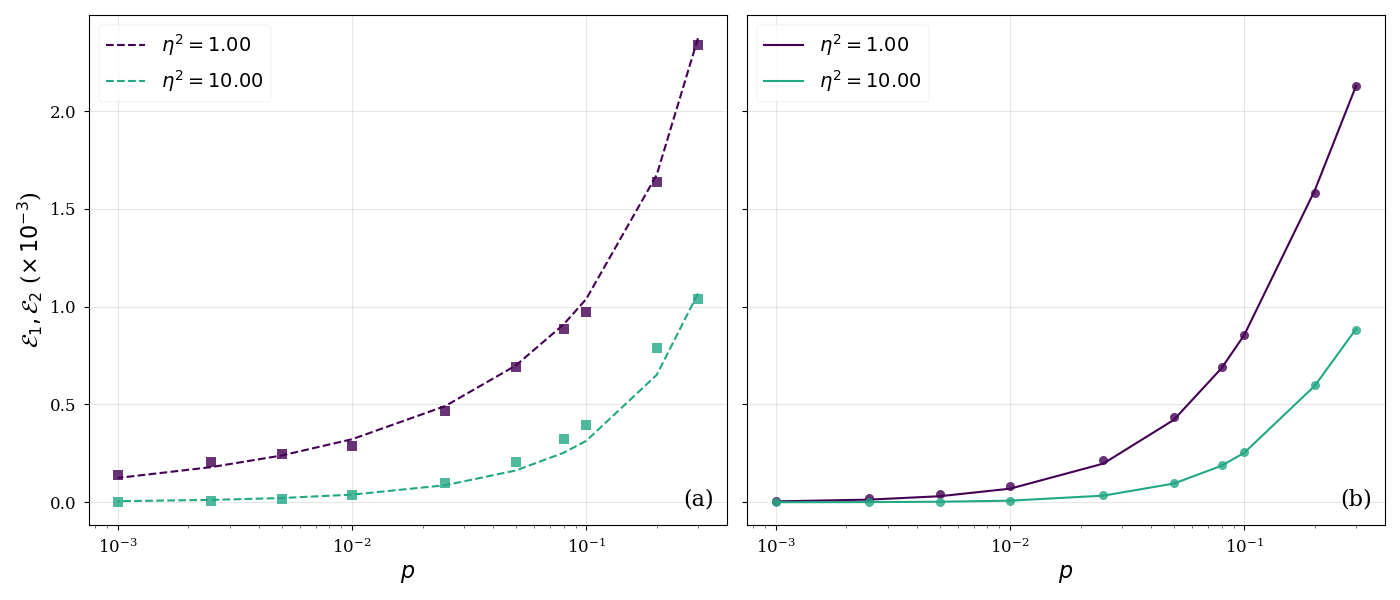}%{Figs/Fig_Model_1_bis.pdf}
	\caption{\footnotesize Empirical and theoretical integrated prediction errors for $M_1\in\mathcal M_1$ and $M_2\in\mathcal M_2$ as functions of the marginal exceedance probability $p$. Panel~(a) shows the error $\mathcal E_1$ of direct classification, represented by squares (\scalebox{0.7}{$\blacksquare$}) and dashed lines. Panel~(b) shows the error $\mathcal E_2$ of distributional regression, represented by circles ($\bullet$) and solid lines. Dark violet and green correspond, respectively, to $\eta^2=1$ and $\eta^2=10$, with $s^2+\sigma^2=2$ in both cases. Symbols are empirical variance estimates over 30 repeated training samples. Dashed and solid lines are the analytical approximations in respectively Eqs.~\eqref{eq:error_M1_finite_main} (for $M_1$) and \eqref{eq:error_M2_finite_main} (for $M_2$) with one model-dependent positive prefactor fitted separately for each method and each value of $\eta^2$. }
	\label{fig:M1M2_pred_errors}
    \vspace*{0cm}
\end{figure*}

Figure~\ref{fig:M1M2_pred_errors} presents empirical estimates of the prediction errors of models $M_1$ and $M_2$ for exceedance probabilities ranging from $p=10^{-3}$ to $p=3\times10^{-1}$. Panel~(a) displays $\mathcal E_1$ for direct classification, represented by squares (\scalebox{0.7}{$\blacksquare$}), whereas panel~(b) displays $\mathcal E_2$ for distributional regression, represented by circles ($\bullet$). The empirical quantities correspond to the variance components defined in Eqs.~\eqref{eq:e1_1} and \eqref{eq:e2_1}. Using the repeated training samples, we also estimated the integrated squared bias from the difference between the mean prediction and the true conditional probability $p_t$. For the configurations reported here, the integrated squared-bias contribution was approximately 5--10 times smaller than the integrated prediction variance, supporting the variance-dominated approximation used in Appendix~\ref{app:asympt_error}. We consider the two noise-to-signal ratios $\eta^2=1$ and $\eta^2=10$, while keeping the marginal variance $\sigma_Y^2=s^2+\sigma^2=2$ fixed. We observe that, at fixed $p$, empirical errors are smaller for the largest value of $\eta^2$. This behavior is consistent with the decreasing intrinsic variability of the conditional probability. Indeed, as discussed in previous section, when $\eta^2\to\infty$, $p_t$ approaches the climatological probability $p$, both ${\cal E}_1$ and ${\cal E}_2$ errors are expected to decrease and the conditional probability therefore becomes easier to predict.

\begin{figure*}[h]
	\centering
	\includegraphics[width=\linewidth]{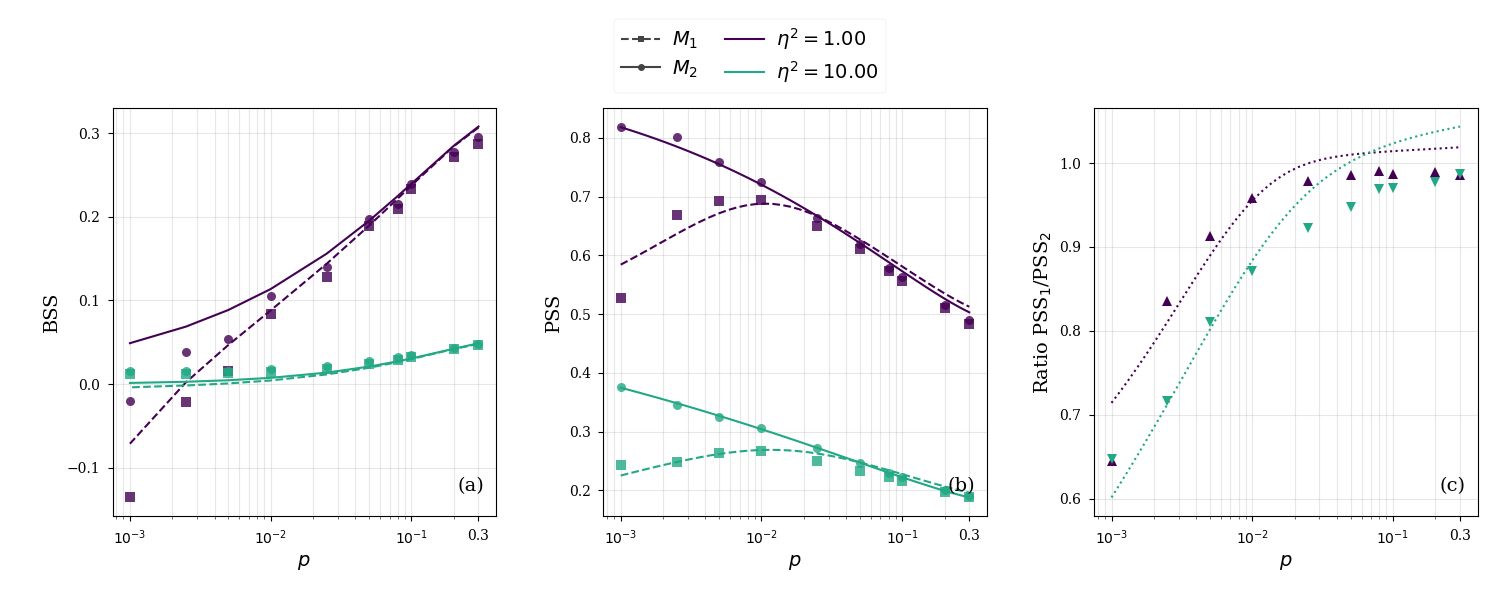}%{Figs/Fig_Model2_4.pdf}
	
	\caption{\footnotesize Comparison of the BSS and PSS for $M_1\in\mathcal{M}_1$ ((\scalebox{0.7}{$\blacksquare$}) and dashed lines) and $M_2\in\mathcal{M}_2$ (($\bullet$) and solid lines). Dark violet and green correspond to $\eta^2=1$ and $\eta^2=10$, respectively. Panel~(a) shows the empirical BSS defined in Eq.~\eqref{eq:def_BSS}, panel~(b) shows the PSS defined in Eq.~\eqref{eq:PSS}, and panel~(c) shows the ratio $\mathrm{PSS}_1/\mathrm{PSS}_2$, all as functions of the marginal exceedance probability $p$. The lines show the theoretical approximations derived in Appendix~\ref{app:logscore} for the BSS, using Eq.~\eqref{BSS_k_vs_E_k}, and in Appendix~\ref{app:pss} for the PSS, using Eqs.~\eqref{eq:pss1_a} and \eqref{eq:pss2_a}. Model-dependent local-variance parameters are fitted separately for each method and each value of $\eta^2$.}
	
	\label{fig:BSS_PSS_xp}
    \vspace*{0cm}
\end{figure*}

The curves in Fig.~\ref{fig:M1M2_pred_errors} show the fitted finite-$p$ analytical approximations in Eqs.~\eqref{eq:error_M1_finite_main} and \eqref{eq:error_M2_finite_main}. Their rare-event limits correspond to Eqs.~\eqref{eq:L2_error_M1} and \eqref{eq:L2_error_M2}, respectively. These expressions are derived in Appendix~\ref{app:asympt_error} under the simplifying assumption that the conditional local variance functions can be approximated by positive constants. For $M_1$, the finite-$p$ curve additionally uses the rare-event simplification $\Delta(z_p;r)\simeq\Phi_2(z_p;r)$ in the numerator. The constants $V_1$ and $V_2$ depend on the model architecture, the full parameter covariance, and the local output sensitivity. They cannot be determined from the scalar Gaussian generative model alone and are therefore fitted separately for each method and each value of $\eta^2$. Consequently, Fig.~\ref{fig:M1M2_pred_errors} primarily tests the predicted dependence of the errors on $p$, rather than providing a parameter-free prediction of their absolute amplitudes. The overall agreement between the fitted curves and the empirical estimates supports the proposed finite-$p$ functional forms and their associated rare-event limits over the probability range considered, while the absolute prefactors remain model-dependent.

These findings are complemented by the skill-score analysis in Fig.~\ref{fig:BSS_PSS_xp}, which compares $M_1$ and $M_2$ using the BSS defined in Eq.~\eqref{eq:def_BSS} and PSS defined in Eq.~\eqref{eq:PSS}. The BSS evaluates probabilistic performance relative to climatology, whereas the PSS evaluates the discrimination of the associated deterministic exceedance predictions.

Panel~(a) shows that the BSS decreases as the noise-to-signal ratio $\eta^2$ increases, although the absolute prediction errors in Fig.~\ref{fig:M1M2_pred_errors} decrease. These results are not contradictory. For large $\eta^2$, $p_t$ becomes less variable and approaches the constant marginal probability $p$. Estimating $p_t$ in absolute terms then becomes easier, but the predictable information available to discriminate the event from climatology decreases, resulting in a lower BSS. Conversely, when $\eta^2$ is small, the distribution of $p_t$ is more dispersed and frequently takes values close to zero or one, making the binary event $I_{t+h}$ more predictable from the covariates. The BSS also generally increases with $p$. Hence, despite the increase of the absolute errors $\mathcal E_1$ and $\mathcal E_2$ with $p$, performance relative to climatology deteriorates for rarer events. This behavior follows from Eq.~\eqref{BSS_k_vs_E_k}, where the estimation error enters through $\mathcal E_k/[p(1-p)]$. For $M_2$, this contribution remains limited in the rare-event regime. For $M_1$, however, Eq.~\eqref{eq:L2_error_M1} gives, up to logarithmic corrections, 
$ \frac{\mathcal E_1}{p} \propto p^{-\frac{2}{2+\eta^2}}$
which explains the pronounced deterioration of $\mathrm{BSS}_1$ at the smallest probabilities, particularly for $\eta^2=1$.

In Fig. \ref{fig:BSS_PSS_xp}(b), we see that very much like BSS, PSS decreases with the noise-to-signal ratio $\eta^2$. Again, this stems from a better quality of $I_t$ prediction for smaller $\eta^2$. The figure further demonstrates that as the exceedance probability $p$ decreases, predictions from model $M_2$ become increasingly accurate, as evidenced by the monotonic increase in $\mathrm{PSS}_2$. This trend confirms our theoretical analysis presented in Section~\ref{ss:analytical} following Eq.~(\ref{eq:pss_2_th}). A similar pattern is observed for $M_1$ at moderate values of $p$. However, $\mathrm{PSS}_1$ reaches a maximum and then decreases for the rarest events. This non-monotonic behavior is consistent with the numerical evaluation of Eq.~\eqref{eq:pss1_a}, which accounts for the loss of effective Bernoulli information in the rare-event regime.

Panel~(c) summarizes the relative performance through $\mathrm{PSS}_1/\mathrm{PSS}_2$. The empirical ratio remains below one over the entire range considered, indicating that $M_2$ consistently outperforms $M_1$ according to the PSS. The difference is relatively small for the most common events but becomes more pronounced as $p$ decreases.

\subsection{Summary and interpretation of the theoretical results}
In summary, the simplified framework introduced in this section allows us to characterize explicitly the information loss induced by thresholding the predictand in approach $\mathcal M_1$, rather than using all continuous observations as in approach $\mathcal M_2$. In the Gaussian location model considered here, the probability estimation errors satisfy Eq.~\eqref{eq:error_ratio_main}, showing that the error of $M_2$ becomes asymptotically smaller than that of $M_1$ as events become increasingly rare. This difference is reflected in the verification scores. In the present model, as $p$ decreases, the exceedance threshold increases and events become increasingly associated with large values of the predictable component $\mu_t$, rather than with extreme realizations of the residual noise alone. Events and non-events therefore become more discriminable from the covariates, causing the PSS based on the true conditional probability to increase. Model $M_2$ can exploit this increasing intrinsic discrimination, whereas, at a fixed training-sample size, the performance of $M_1$ eventually deteriorates for the rarest events because threshold-specific positive labels become increasingly scarce. The BSS, which measures probabilistic performance relative to climatology, instead decreases toward rarer events for both approaches, indicating that maintaining probabilistic skill relative to the climatological benchmark becomes increasingly difficult. Nevertheless, ${M}_2$ retains a clear advantage over ${M}_1$ in this regime. 
The logarithmic score, which is also a strictly proper scoring rule, leads to the same relative ranking of the two approaches. 
The noise-to-signal ratio $\eta^2$ controls a distinct aspect of the problem, namely the intrinsic predictability of the exceedance event. In this respect, increasing $\eta^2$ may be interpreted qualitatively as representing the loss of predictability associated with an increasing forecast horizon. At fixed, sufficiently small $p$, the probability-estimation errors decrease with increasing $\eta^2$ in the large-noise regime. This reflects the fact that the conditional probability $p_t$ fluctuates less around the climatological probability $p$ and therefore becomes easier to estimate in absolute terms. At the same time, $p_t$ carries progressively less covariate-dependent information for distinguishing events from non-events. Consequently, the BSS and PSS decrease as $\eta^2$ increases and vanish in the pure-noise limit.

Overall, this tractable framework does not provide universal rare-event scaling laws, since the specific dependence on $p$ and $\eta$ follows from the Gaussian location model in Eq.~\eqref{eq:observation_model} and its estimation assumptions, including a known and constant conditional variance. Its main purpose is to isolate the general information loss induced by predictand thresholding. A similar qualitative advantage for distributional modeling may arise when exceedance probabilities are driven by predictable changes in location. Approach $\mathcal M_2$ can infer such changes from the continuous response, whereas $\mathcal M_1$ uses only threshold-specific binary labels. The next section investigates whether this qualitative mechanism remains visible for real meteorological variables under non-Gaussian, heteroskedastic and potentially misspecified conditions.

%%%%%%%%%%%%%%%%%%%%%%%%%%%%%%%%%%%%%%%%%%%%%%%%%%%%%%%%%%%%%%%%%%%%%%%%%%%%
%%%% APPLICATION TO WIND AND RAINFALL FORECAST
%%%%%%%%%%%%%%%%%%%%%%%%%%%%%%%%%%%%%%%%%%%%%%%%%%%%%%%%%%%%%%%%%%%%%%%%%%%%
\section{Application to rainfall and wind speed data}
\label{sec:application}
%%%%%%%%%%%%%%%%%%%%%%%%%%%%%%%%%%%%%%%%%%%%%%%%%%%%%%%%%%%%%%%%%%%%%%%%%%%%

In this section, we compare the two forecasting approaches for strong surface winds and intense hourly accumulated rainfall using ground-station observations and NWP forecasts. To ensure a direct comparison, $M_1$ and $M_2$ use the same predictors and hybrid neural-network backbone, differing only in their output representation and training objective. We first describe the data and forecasting tasks, followed by the model architecture, training procedure and empirical results. 

These applications differ substantially from the idealized setting of Section~\ref{sec:theormod}. Wind speed and rainfall have non-Gaussian distributions, their conditional dispersion and shape may vary with the predictors and the selected parametric families may be misspecified. The quantitative scaling laws derived from the Gaussian location model are therefore not expected to apply directly. Instead, the applications examine whether its main qualitative implication remains visible under more realistic conditions, namely whether the relative advantage of distributional modeling increases as the exceedance probability decreases.

%%%% Meteonet Dataset
\subsection{The MeteoNet dataset}

\begin{figure}[h!]
	\centering
	\includegraphics[width=0.95\linewidth]{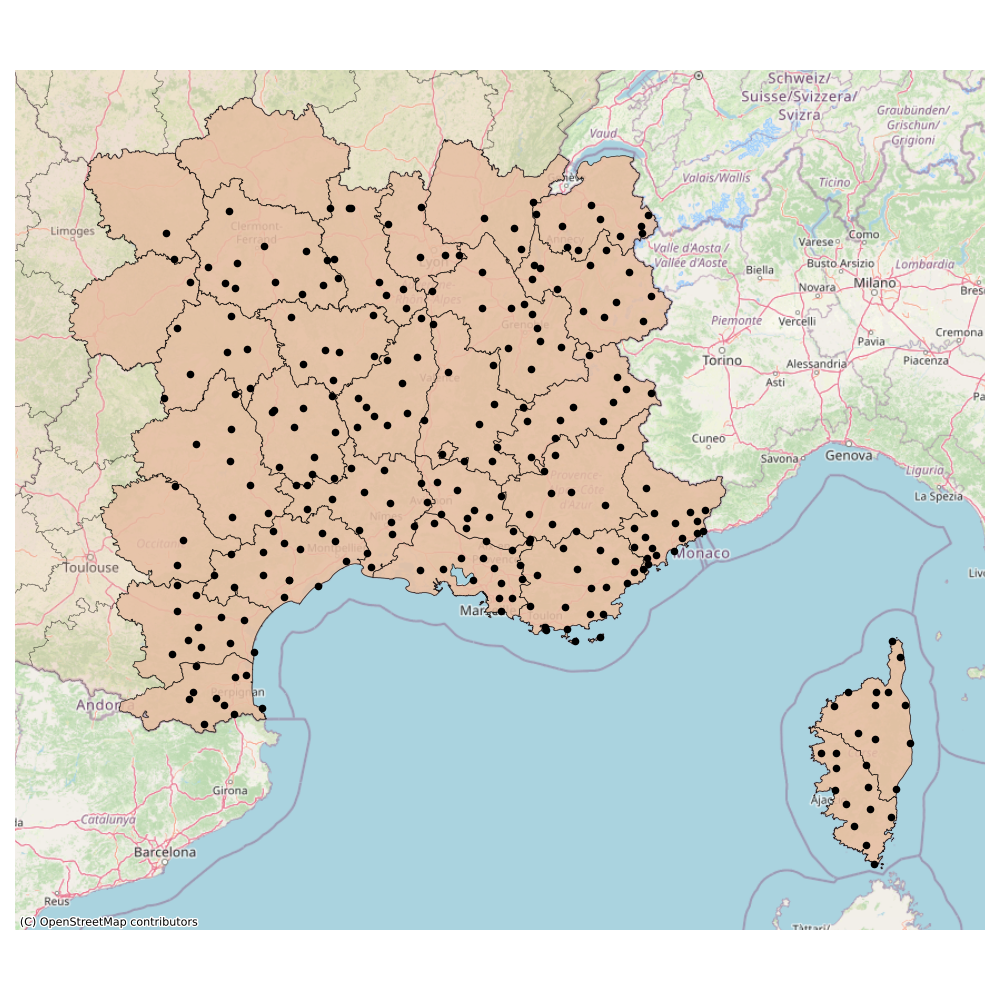}%{Figs/map_stations.png}
	\vspace*{-0.5cm}
	\caption{\footnotesize Geographical extent of the southeastern MeteoNet domain and locations of the 278 ground stations retained for the wind-speed analysis ($\bullet$).}
	\label{fig:sta_loc}
\end{figure}
The meteorological data used in this study are obtained from MeteoNet, an open dataset provided by Météo-France for machine-learning research \citep{larvor2021meteonet}. MeteoNet covers southeastern and northwestern France over the period 2016--2018 and combines ground-station observations with gridded NWP forecasts. Following \citet{baggio2025local}, we focus on the southeastern domain and use ground-station observations together with AROME and ARPEGE forecasts.
The raw data are processed using the two-stage procedure described in detail by \citet{baggio2025local}. First, the observations and NWP fields are reorganized into station-centered subsets containing the local station time series, observations from neighboring stations and local patches extracted from the two- and three-dimensional NWP fields. Second, these subsets are converted into model-ready samples. Samples containing missing or invalid values are discarded 
and the final features are normalized and exported in a format compatible with PyTorch data generators.
Stations are retained only when the selected variables contain a sufficient number of valid observations. We require at least 2000 valid hourly samples for the wind-speed application and 500 for the accumulated-rainfall application. These criteria leave 278 stations for wind speed and 268 stations for rainfall among those available in the original dataset. Figure~\ref{fig:sta_loc} shows the stations retained for the wind-speed analysis.

The input variables used for each predictand are summarized in Table~\ref{tab:input_data}. Ground-station observations, originally available at a 6-min resolution, are aggregated to hourly values. Station-based inputs include the station of interest and its ten nearest neighbors, together with the current and six previous hourly observations. For the NWP inputs, we use surface fields from the high-resolution AROME model at $0.025^\circ$ resolution and three-dimensional fields from ARPEGE at $0.1^\circ$ resolution. Local patches centered on each station contain $11\times11$ horizontal grid points for AROME and $5\times5$ grid points at seven atmospheric levels for ARPEGE, resulting in a $7\times5\times5$ tensor for each three-dimensional variable and forecast time. For each day, the NWP predictors are extracted from the 00:00~UTC forecast cycle. AROME fields are available hourly, whereas the temporal resolution of the ARPEGE fields varies between 1 and 3~h according to lead time. The construction of the final input tensors is described in Subsection~\ref{subsec:input_desig}. Further implementation details on the MeteoNet preprocessing pipeline are provided by \citet{baggio2025local}.
 
\begin{table*}[t]
\centering
\caption{Summary of the MeteoNet input data used for model training. For NWP inputs, the spatial grid denotes a local horizontal subdomain centered on the ground station of interest. For ARPEGE, the additional dimension corresponds to the atmospheric levels retained in the input tensor.}\label{tab:input_data}
\small
\setlength{\tabcolsep}{4pt}
\begin{tabular}{llll}
\hline
Input type & Space and time grids & Wind speed target & Accumulated-rainfall target \\
\hline
Stations &
\begin{tabular}[t]{@{}l@{}}
Spatial grid: 11 locations\\
(station + 10 neighbours)\\
Time grid: current + 6 past hourly values
\end{tabular} &
\begin{tabular}[t]{@{}l@{}}
Wind components $u$, $v$ (m\,s$^{-1}$)\\
Temperature (K)
\end{tabular} &
\begin{tabular}[t]{@{}l@{}}
Wind components $u$, $v$ (m\,s$^{-1}$)\\
Temperature (K)\\
Relative humidity (\%)\\
Precipitation (mm\,h$^{-1}$)
\end{tabular} \\
\hline
AROME &
\begin{tabular}[t]{@{}l@{}}
Spatial grid: $11 \times 11$\\
(centered on the station)\\
Time grid: all 6\,h-ahead predictions
\end{tabular} &
\begin{tabular}[t]{@{}l@{}}
2\,m temperature (K)\\
2\,m relative humidity (\%)\\
Wind components $u$, $v$ (m\,s$^{-1}$)\\
MSLP (Pa)
\end{tabular} &
\begin{tabular}[t]{@{}l@{}}
Same as wind-speed target,\\
plus 2\,m dew-point temperature (K)\\
and total precipitation (mm)
\end{tabular} \\
\hline
ARPEGE &
\begin{tabular}[t]{@{}l@{}}
Spatial grid: $7 \times 5 \times 5$\\
(centered on the station)\\
Time grid: all 6\,h-ahead predictions
\end{tabular} &
\begin{tabular}[t]{@{}l@{}}
Temperature (K)\\
Wind components $u$, $v$ (m\,s$^{-1}$)\\
Pressure (Pa)
\end{tabular} &
\begin{tabular}[t]{@{}l@{}}
Same as wind-speed target,\\
plus vertical velocity (Pa\,s$^{-1}$)
\end{tabular} \\
\hline
\end{tabular}

\end{table*}

%%%%%%%%%%%%%%%%%%%%%%%%%%%%%%%%%%%%%%%%%%%%%%%%%%%%%%%%%%%%%%%%%%%%%%%%%%%%
%%%% Forecasting Task
\subsection{Statement of the forecasting problem for wind and cumulative rainfall}
We now adapt the general forecasting problem formulated in Section~\ref{sec:extremebins} to the wind-speed and accumulated-rainfall applications. For a weather variable $Y_t$ observed at station $S$, we consider threshold-exceedance forecasts at the lead times $\mathbf h=(h_1,\ldots,h_H)$. At forecast initialization time $t$, the corresponding vector of future binary outcomes is 
\begin{equation} 
\mathbf I_{S,t+\mathbf h} = \begin{pmatrix} I_{S,t+h_1}\\ I_{S,t+h_2}\\ \vdots\\ I_{S,t+h_H} 
\end{pmatrix}, 
\end{equation} 
where each component indicates whether the selected threshold is exceeded at station $S$ and lead time $h_j$. Following \citet{baggio2025local}, we set $H=6$ and $\mathbf h=(1,\ldots,6)$~h. Both models simultaneously produce forecasts for these six lead times. For $M_1$, the training objective is obtained by summing the binary cross-entropy losses defined in Eq.~\eqref{eq:bce_loss} over the six horizons. For $M_2$, the negative log-likelihood in Eq.~\eqref{eq:log_likelihood} is summed in the same way. Thus, the models share information through their common neural-network representation, but the loss does not explicitly model the joint dependence among forecast errors at different horizons. Developing a joint multihorizon probabilistic model that represents these dependencies is left for future work.

For the two weather variables we consider, namely  hourly wind speed (m/s) and 1-hour accumulated rainfall (mm), thresholds $Q_p$ are calculated \emph{station-wise}, that is, in Eq.~(\ref{eq:def_quantile}), the climatological CDF are station-specific: $F_C(y)=F_C^S(y)$. For wind speed, the quantiles $Q_p$ are computed for the following probabilities $p$: 
\begin{equation}
    p \in \left \{ 0.2,0.1,0.08,0.05,0.03,0.01,0.005,0.002\right \}\,.
    \label{eq:thresholds_wind}
\end{equation}
For rainfall, the large proportion of dry days motivates the use of quantiles conditional on rainfall occurrence.
Let $F_{C,+}^S$ denote the station-wise climatological CDF conditional on rainfall occurrence $F_{C,+}^S(y) = \mathrm{Pr}(Y \le y \mid Y > 0)$. The quantiles $Q_{p,S}$ are defined with respect to this conditional distribution,  i.e. $F_{C,+}^S(Q_{p,S}) = 1-p_{+}$, where $p_{+}$ is such that: 
\begin{equation}
    p_{+} \in \left\{0.5,0.4,0.3,0.25,0.2,0.15,0.1,0.05,0.025\right \} \,.
        \label{eq:thresholds_rain}
\end{equation}
%The subscript $R$ distinguishes rainfall from wind speed, denoted by $W$. 
When displaying results, metrics are plotted as a function of the ``unconditionnal'' $p$, defined as $p = \bar{p}_{\mathrm{wet}} \, p_{+}$, where $p_{\mathrm{wet}} = \mathrm{Pr}(Y>0)$ is the wet-event probability at a single site and $\bar{p}_{\mathrm{wet}}$ denotes its average across all considered stations. For the present dataset $\bar{p}_{\mathrm{wet}} \approx 0.08$. Because $p_{\mathrm{wet}}$ varies across stations, $p$ is an approximate common unconditional probability rather than the exact station-specific probability. This convention nevertheless facilitates comparison with the theoretical results, which are expressed in terms of the marginal exceedance probability.

%%%% more details on the probabilistic modelling and corresponding output layer   
\subsubsection{Probabilistic models for surface wind speed and rainfalls}
When using a probabilistic model of type $\mathcal{M}_2$, different parametric forms for the implied conditional density function $f(y)$ are adopted to model wind speed and accumulated rainfall. It is worth emphasizing that, although these distributions have been selected with care, the differences among alternative parametric families remain limited once fundamental physical constraints of the target variable are properly enforced, as discussed later in Subsection~\ref{subsec:modelspec}.
%%%% probabilistic modelling of wind speed   
\paragraph{Wind speed}
For wind speed, we adopt the so-called \emph{Multifractal-Rice} (M-Rice) distribution, following previous work in \citet{BaggioMuzy2024}, where it was shown to outperform classical alternatives such as the Weibull and Gamma distributions in forecasting applications.  The M-Rice distribution, introduced in \citet{BaMuPo11}, is motivated by the random cascade framework used to describe fully developed turbulence. It generalizes the classical Rice distribution by allowing its scale parameter to be random, typically modeled as a log-normal variable, thereby incorporating intermittency effects. The resulting distribution is characterized by three parameters $(\nu, \sigma^2, \lambda^2)$. The ``non-centrality'' parameter is the Euclidean norm of the mean vector $(\mu_1,\mu_2)$ ($\nu=\sqrt{\mu_1^2+\mu_2^2}$) of the underlying bivariate conditional Gaussian distribution. $\sigma^2$ governs dispersion and $\lambda^2$, often referred to as the \emph{intermittency parameter}, mainly regulates the tail behavior. In particular, larger values of $\lambda^2$ produce heavier tails, increasing the probability assigned to extreme wind speeds. 
The formal definition of the M-Rice distribution, together with a detailed interpretation of its parameters and numerical implementation, is provided in Appendix~\ref{app:M-Rice}.
%%%% probabilistic modelling of accumulated rainfall   
\paragraph{Accumulated rainfall} 
For hourly accumulated rainfall, we adopt a mixed lognormal distribution to account for the discrete-continuous nature of precipitation. Rainfall observations are characterized by a substantial probability mass at zero, corresponding to dry conditions, together with a positively skewed continuous distribution for positive amounts. The mixed lognormal model represents this structure by combining a point mass at zero with a lognormal distribution for strictly positive values \citep{cho2004comparison,kedem1990estimation}. The distribution has three parameters: the rainfall-occurrence probability $p_{\mathrm{wet}}$ and the lognormal parameters $(\mu,\sigma^2)$ governing the location and dispersion of positive rainfall amounts. Its mathematical formulation is provided in Appendix~\ref{app:mixture_rainfall}. We also tested alternative mixed distributions, which produced similar results. These distributions are also defined in the Appendix~\ref{app:mixture_rainfall}.  

%%%%%%%%%%%%%%%%%%%%%%%%%%%%%%%%%%%%%%%%%%%%%%%%%%%%%%%%%%%%%%%%%%%%%%%%%%%%
%%%% Neural Network
\subsection{Hybrid neural network for predicting cumulative precipitation or surface wind speed}
The forecasting models use the hybrid neural-network architecture introduced in \citet{baggio2025local} and reimplemented in PyTorch. The architecture combines heterogeneous station and NWP inputs through dedicated processing branches. This subsection describes the input tensors, model architecture and training procedure.
%%%% Input design
\subsubsection{Input design}
\label{subsec:input_desig}
The heterogeneous data sources described above are combined into the ANN model inputs with each data type processed by a dedicated branch (see Subsection~\ref{subsec:modarch}). A feature--target pair $(\textbf{{X}}_\texttt{k},\textbf{{Y}}_\texttt{k})$ 
is constructed for each sample index, hereafter referred to as a \emph{key}, $\mathrm{k}=(S,d,t)$, where $S$ denotes the station, $d$ the date, and $t$ the initialization hour.
The target $\mathbf Y_{\mathrm{k}}$ contains the six future values of either wind speed or accumulated rainfall, corresponding to forecast horizons from 1 to 6~h. The input is written as $\textbf{{X}}_\texttt{k} = \Big[\textbf{{GS}}_\texttt{k}, \textbf{{AR}}_\texttt{k}, \textbf{{AP}}_\texttt{k}, \textbf{{C}}_\texttt{k}, \textbf{{D}}_\texttt{k} \Big],$, where  $\textbf{{GS}}_\texttt{k}$, $\textbf{{AR}}_\texttt{k}$, and $\textbf{{AP}}_\texttt{k}$ contain ground-station, AROME, and ARPEGE features, respectively, while $\textbf{{C}}_\texttt{k}$ and $\textbf{{D}}_\texttt{k}$ contain temporal, geographical, and station-neighborhood information. The ground-station tensor $\textbf{{GS}}_\texttt{k}$ contains observations from the target station $S$ and its 10 nearest neighbors. Let $n_S$ denote the number of variables retained at each station. The input dimension at each time step is therefore $11n_S$, corresponding to 33 variables for wind speed and 55 for rainfall, as reported in Table~\ref{tab:input_data}. Using the current observation and the six preceding hourly observations gives tensors of shape $7\times33$ for wind speed and $7\times55$ for rainfall. The AROME tensor $\textbf{AR}_{\mathrm{k}}$ is extracted from an $11\times11$ patch centered on station $S$, corresponding to a spatial extent of approximately $0.3^\circ$. Forecasts at lead times from 1 to 6~h are retained, giving tensors of shape $6\times n_{\mathrm{AR}}\times11\times11$, where $n_{\mathrm{AR}}=5$ for wind speed and $n_{\mathrm{AR}}=7$ for rainfall. The ARPEGE tensor $\textbf{AP}_{\mathrm{k}}$ is constructed similarly from a $5\times5$ horizontal patch centered on station $S$, corresponding to an approximate spatial extent of $0.5^\circ$. As previously mentionned, the selected variables are extracted at seven atmospheric levels and forecast horizons are matched to the closest available times when hourly fields are unavailable. The resulting tensors have shape $6\times4\times7\times5\times5$ for wind speed and $6\times5\times7\times5\times5$ for rainfall. Finally, $\textbf C_{\mathrm{k}}$ contains cyclic encodings of hour and day together with the latitude, longitude and altitude of the target station, while $\textbf D_{\mathrm{k}}$ contains the relative positions of its ten neighboring stations. Further implementation details are provided in \citet{baggio2025local}.

\subsubsection{Neural network model architecture}\label{subsec:modarch}
The proposed ANN architecture follows the design introduced in \citet{baggio2025local}, with minor adaptations. Each component of the input $\textbf{{X}}_{\mathrm{k}}$ is processed by a dedicated branch tailored to its structure. The ground-station time series $\textbf{{GS}}_{\mathrm{k}}$ are processed by stacked LSTM layers to capture temporal dependencies. The spatiotemporal AROME and ARPEGE tensors,  $\textbf{{AR}}_\texttt{k}$ and  $\textbf{{AP}}_\texttt{k}$, are first processed by convolutional layers to extract spatial features and subsequently passed to LSTM layers to encode their temporal evolution. 
In each branch, contextual features are encoded by fully connected layers and concatenated with the other predictors before the recurrent layers. The outputs of the three branches are subsequently concatenated and processed by an additional dense block before the final output layers. Further architectural details are provided in \citet{baggio2025local}. 
For the classification model $M_1$, sigmoid activations are applied to the final outputs to produce exceedance probabilities for the six forecast lead times. For the distributional model $M_2$, the final dense representation is mapped, through separate output layers, to the parameters of the selected predictive distribution at each lead time. Appropriate activation functions are used to enforce the constraints associated with each distributional parameter.

%%%% dataset split
\subsubsection{Dataset split} Splitting time-series data requires limiting information leakage between temporally dependent samples while maintaining representative seasonal coverage across the subsets \citep{Schultz2021}. Given the limited three-year period available here (2016--2018), we adopted a date-based random split. All samples, forecasting tasks and targets associated with the same calendar date were assigned to the same subset, thereby preventing the most strongly overlapping samples from being distributed across different subsets. In addition, forecasts initial times were restricted to times up to 17:00~UTC or earlier, ensuring that all six forecast targets remained within the same calendar date. Together with the date-based assignment, this restriction provides a natural temporal separation between consecutive days that mitigates leakage while preserving seasonal variability. The available dates were randomly assigned to the training, validation, and test subsets with respective proportions of $85\%$, $10\%$, and $5\%$. Samples were then constructed for each subset using only dates for which the required observations and NWP inputs were available and passed the quality-control criteria described above. With the input configuration reported in Table~\ref{tab:input_data}, the resulting datasets contain approximately four million sample keys. The wind-speed dataset contains approximately $3.52\times10^6$, $3.93\times10^5$, and $2.23\times10^5$ training, validation and test keys, respectively. The corresponding rainfall subsets contain $3.39\times10^6$, $3.78\times10^5$ and $2.15\times10^5$ keys.

\subsubsection{Hyperparameter selection and model training}
%%%CITE NUMBER OF PARAMETER
The network presented above contains a total of around $2.7 \times 10^5$ trainable parameters, which is relatively small by modern standards. The model was trained using the Adam optimizer. To mitigate potential overfitting, we adopted an early stopping strategy based on the validation loss (with a patience parameter of $10$ for wind speed, $20$ for accumulated rainfall). The remaining hyperparameters were selected based on prior experience and are reported in Table~\ref{tab:hyperparameters}.
\begin{table}[h]
	\caption{\footnotesize training hyperparameters used in the ANN.}
	\label{tab:hyperparameters}
\centering
\begin{tabular}{lc}

\hline
\textbf{Training hyperparameters} & \\
\hline
Learning rate & 0.001 \\
LSTM dropout rate & 0.02 \\
%Number of epochs & 250 \\
Batch size (training) & 512 \\
Early stopping patience & 10 (wind), 20 (rainfalls) \\
%Random seed & 3 \\
\hline
\end{tabular}
\end{table}
%%%AVERAGE TRAINING TIME
With this setup, model training takes about 3 to 4 minutes per epoch on a single NVIDIA Tesla V100 GPU. Considering that the very first epoch takes approximately twice this time due to initialization overhead and that early stopping occurs after 11--25 epochs, training takes less than 2 hours per model. It is important to emphasize a fundamental difference between the two approaches: while $M_2$, which models the full predictive distribution, relies on a single model for all considered thresholds (Eqs.~(\ref{eq:thresholds_wind}),~(\ref{eq:thresholds_rain})), strategy ${M}_1$, based on binary classification, requires training a separate model for each threshold. After training, inference from the ANN is very fast, so that forecasts can be issued in a matter of seconds (see \citet{baggio2025local} for details). Since we did not perform an extensive optimization of the hyperparameter space, the model parameters were kept fixed at standard baseline values, leading to highly consistent results between the validation and test splits. Accordingly, the scores presented in Subsection \ref{sub:results} are computed over both datasets simultaneously. This choice is motivated by our primary interest in the structural behavior of the curves as $p \to 0$, rather than in the absolute metric values themselves. In practice, evaluating solely on the test set produces noisier estimates because of its smaller sample size, while the relative ranking of the models remains unchanged across both subsets.
%###RESULTS
\subsection{Application results}\label{sub:results}
In this Section, we compare the two modeling strategies $\mathcal{M}_1$ and $\mathcal{M}_2$ and interpret their relative performance in light of the qualitative mechanism introduced in Section~\ref{sec:theormod}. Because the observation distributions and models differ from the Gaussian toy problem, the empirical curves are not expected to reproduce the theoretical powers of $p$ quantitatively. The relevant theoretical signature is instead an increasing relative advantage of $\mathcal{M}_2$ as the exceedance probability decreases. We report results for forecast horizons $h=1$ and $h=6$; intermediate horizons behave similarly and generally lie between these two cases. Increasing the forecast horizon is expected to reduce the predictable signal relative to the unresolved variability, qualitatively resembling an increase in the noise-to-signal ratio $\eta^2$ of the toy model. This analogy is used only as an interpretive guide and does not imply that an effective value of $\eta$ is quantitatively estimated from the applications.

\subsubsection{Hourly wind speed} 
\begin{figure*}[ht]
\centering
\includegraphics[width=\linewidth]{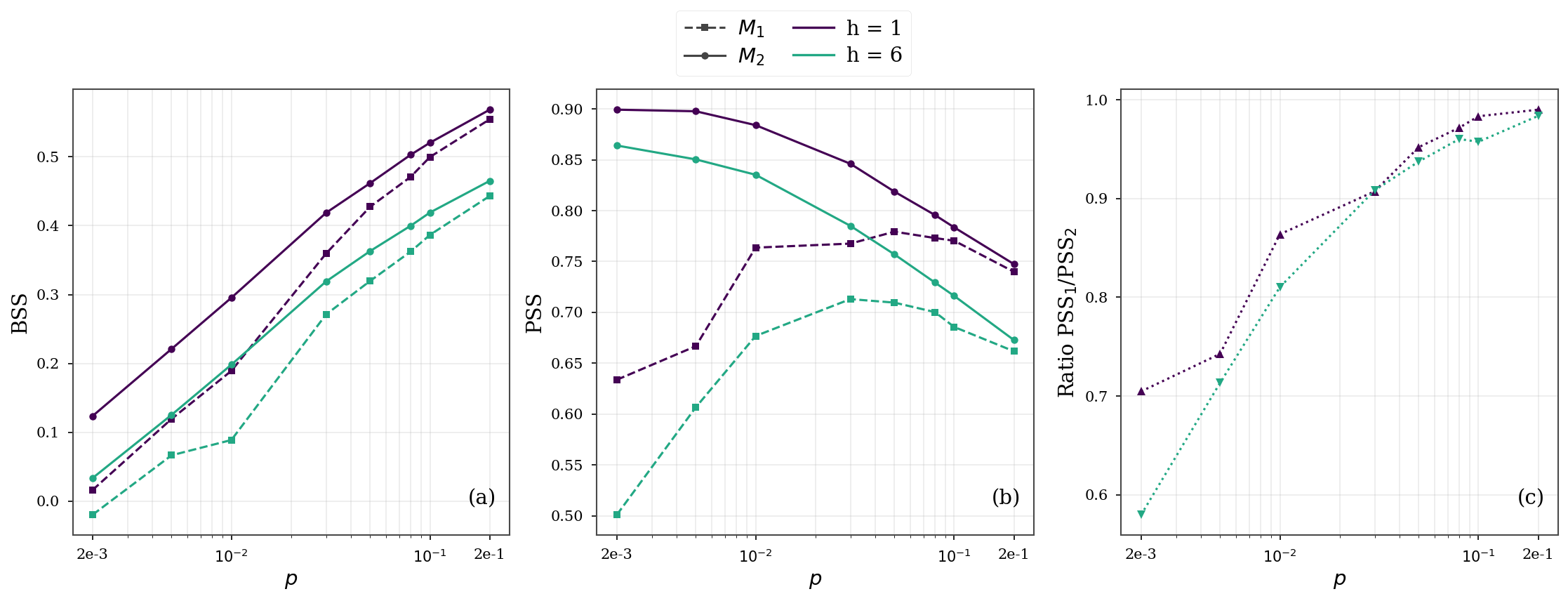}%{Figs/WIND_results_PSS_BSS.png}
\caption{\footnotesize $\text{BSS}$ (panel ~(a)), $\text{PSS}$ (panel ~(b)) and the ratio $\mathrm{PSS}_1/\mathrm{PSS}_2$ (panel ~(c)) for \textbf{ hourly wind speed}. Results are shown for the two models ${M}_1$ (symbols (\scalebox{0.7}{$\blacksquare$}) and dashed lines) and ${M}_2$ (symbols ($\bullet$) and solid lines) at two forecast horizons, $h=1$ h (violet) and $h=6$ h (green).}
\label{fig:Wind_PSS_BSS}
\vspace*{0cm}
\end{figure*}

%%%PSS and BSS
In Fig.~\ref{fig:Wind_PSS_BSS} values of $\text{BSS}$, $\text{PSS}$ and the $\text{PSS}$ ratio are shown. In line with what observed for the Gaussian location model, metrics associated with model ${M}_2$ are  consistently better than the ones obtained with the classification approach ${M}_1$. Moreover, the overall trend of  $\text{BSS}$ and $\text{PSS}$ for $p\to0$ reflects what shown in Fig.~\ref{fig:BSS_PSS_xp}. 
%%---------------------------------------------------
%%%PSS
In particular $\text{PSS}_2$ increases steadily with decreasing $p$, while $\text{PSS}_1$, though behaving similarly for intermediate values of $p$, sharply deteriorates when $p \to 0$ (Fig.~\ref{fig:Wind_PSS_BSS}(c)). This behaviour is reflected in the trend  displayed by ratio $\text{PSS}_1/\text{PSS}_2$, whose value is near to $1$ for intermediate values of $p$ but decreases as $p \to 0$, thus reflecting what predicted by the model curve (Fig.~\ref{fig:BSS_PSS_xp}(b)).  This means that model ${M}_2$ maintains substantially higher discrimination ability in the rare-event regime, whereas ${M}_1$ exhibits a marked degradation. Similar behaviour is observed at all forecast horizons, though skill degrades with increasing $h$.
%%%BSS
In agreement with what observed for the simplified model, the two approaches are characterized by similar values of BSS for moderate values of $p$ and  a faster degradation of $\text{BSS}_1$ with respect to $\text{BSS}_2$ is observed when $p\to 0$  (Figs ~\ref{fig:Wind_PSS_BSS}(a) and ~\ref{fig:BSS_PSS_xp}(a)). Moreover, when considering the $\text{BSS}$ decomposition ~(\ref{eq:def_BSS}) (not shown) we observed that model ${M}_2$ performs better both in terms of reliability and resolution. 
For both models, the normalized reliability penalty $\frac{Rel}{U}$ increases toward rarer events, but remains smaller than the corresponding resolution contribution $\frac{Res}{U}$ over the considered range. Hence, the gain associated with resolution continues to exceed the penalty due to miscalibration, resulting in positive BSS. The smaller reliability contribution of ${M}_2$ indicates better calibration relative to ${M}_1$, particularly at the rarest thresholds.
\begin{figure*}[ht]
	\centering
	\includegraphics[width=\linewidth]{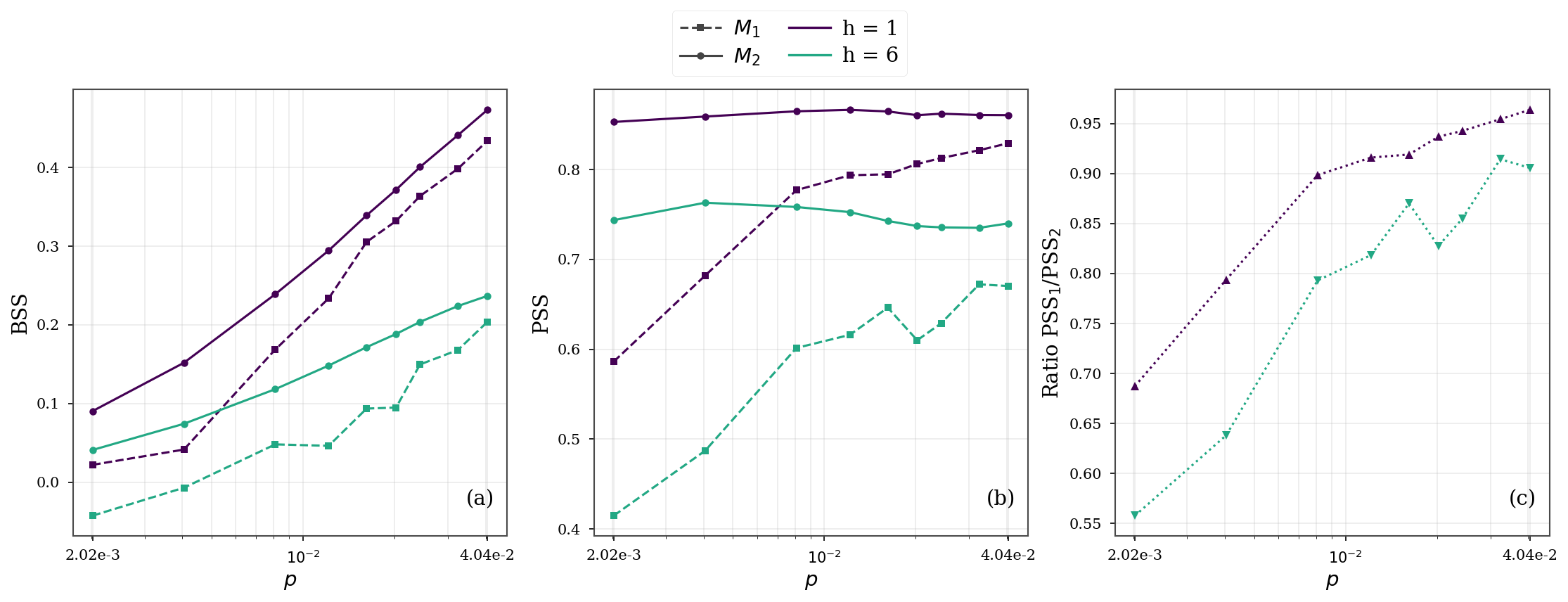}%{Figs/RAIN_results_PSS_BSS.png}
	\caption{ $\text{BSS}$ (panel ~(a)), $\text{PSS}$ (panel ~(b) and the ratio $\mathrm{PSS}_1/\mathrm{PSS}_2$ (panel ~(c)) for  \textbf{hourly accumulated rainfall} are shown for the two models ${M}_1$ (symbols (\scalebox{0.7}{$\blacksquare$}) and dashed lines) and ${M}_2$ (symbols ($\bullet$) and solid lines). Two different forecast horizons are highlighted: $h=1$ h (in violet) and $h=6$ h (green).   }
	\label{fig:RAIN_PSS_BSS}
    \vspace*{0cm}
\end{figure*}
%%---------------------------------------------------
\begin{table}%[t]
\caption{AUC and LS scores for \textbf{hourly wind speed} forecasts at lead times $h=1$ and $6\,\mathrm{h}$, and probabilities $p=0.05$ and $0.005$.}
\label{tab:WIND_AUC_logScore}
\centering
\small
\setlength{\tabcolsep}{6pt}
\begin{tabular}{llllll}
\hline
Model & $p$ & \multicolumn{2}{c}{AUC} & \multicolumn{2}{c}{LS} \\
\cline{3-6}
 &  & $1\,\mathrm{h}$ & $6\,\mathrm{h}$ & $1\,\mathrm{h}$ & $6\,\mathrm{h}$ \\
\hline
\multirow{2}{*}{$\mathcal{M}_1$} 
& 0.05   & 0.958 & 0.934 & 0.095 & 0.129 \\
& 0.005 & 0.963 & 0.955 & 0.019 & 0.024 \\
\multirow{2}{*}{$\mathcal{M}_2$} 
& 0.05   & 0.965 & 0.943 & 0.085 & 0.111 \\
& 0.005 & 0.982 & 0.969 & 0.010 & 0.014 \\
\hline
\end{tabular}
\end{table}
Finally values of the AUC and LS (Eq.~(\ref{eq:def_logScore})) are reported in Table ~\ref{tab:WIND_AUC_logScore} for $p=0.05,\,0.005$. Model $M_2$ outperforms $M_1$ according to both scores and at both horizons. The AUC difference becomes larger at $p=0.005$, while the LS values confirm the same relative ranking of the two approaches.
%\FloatBarrier
%%%%%%%%%%%%%%%%%%%%%%%%%%%%%%%%%%%%%%%%%%%%%%%%%%%%%%%%%%%%%%%%%%%%%%%%%%%%%
%%RAIN

\subsubsection{Hourly accumulated rainfall}	
The analysis presented in the case of hourly wind speed is now  extended to accumulated rainfall.
%%---------------------------------------------------
%%%PSS and BSS
The behaviour of $\text{BSS}$, $\text{PSS}$ and ratio $\mathrm{PSS}_1/\mathrm{PSS}_2$ are displayed in Fig.~\ref{fig:RAIN_PSS_BSS} over the range of unconditional probabilities.
%%%PSS rain
The PSS exhibits a different dependence on event rarity from that observed for wind speed. As $p$ decreases, $\mathrm{PSS}_1$ decreases strongly, whereas $\mathrm{PSS}_2$ remains approximately constant at each forecast horizon (Fig. ~\ref{fig:RAIN_PSS_BSS}(b)). Consequently, the ratio $\mathrm{PSS}_1/\mathrm{PSS}_2$ decreases toward rarer events (Fig. ~\ref{fig:RAIN_PSS_BSS}(c)), showing that the relative advantage of $M_2$ increases primarily because direct classification deteriorates, rather than because the discrimination of $M_2$ improves. This behavior differs from the wind-speed results, for which $\mathrm{PSS}_2$ increases more clearly toward rarer exceedances (Fig. ~\ref{fig:Wind_PSS_BSS}(b)). The rainfall results therefore provide a less direct representation of the idealized location-shift mechanism. They suggest that increasingly intense rainfall events are not associated with progressively more distinctive covariate configurations over the probability range considered. One possible interpretation is that a substantial part of the predictable signal is carried by the conditional rainfall-occurrence probability $p_{\mathrm{wet}}(t)$, rather than by increasingly distinctive precursors of extreme positive rainfall amounts. The predictors may therefore distinguish dry from wet conditions more effectively than moderate from extreme rainfall. Importantly, the absence of an increasing trend in $\mathrm{PSS}_2$ does not imply poor discrimination. At $h=1$, $\mathrm{PSS}_2$ remains close to $0.85$ over the entire probability range, while at $h=6$ it remains approximately between $0.74$ and $0.76$. The distributional model therefore maintains substantial discrimination even though this discrimination does not increase toward the rarest rainfall events.
The BSS decreases toward rarer events for both approaches (Fig.~\ref{fig:RAIN_PSS_BSS}(a)), as was also observed for wind speed. Model $M_2$ consistently outperforms $M_1$ and the difference between the methods becomes more pronounced at the rarest thresholds. At $h=6$, $\mathrm{BSS}_1$ becomes negative for the smallest probability levels, indicating performance below the climatological reference, whereas $\mathrm{BSS}_2$ remains positive. A separate examination of the reliability--resolution decomposition in Eq.~\eqref{eq:def_BSS} leads to conclusions similar to those obtained for wind speed, with $M_2$ showing both a smaller reliability penalty and a larger resolution contribution. A feature that is more clearly visible for rainfall is the dependence of the slope of $\mathrm{BSS}(p)$ on forecast horizon $h$, a behavior that is less apparent for wind speed and is reminiscent of the theoretical model, in which the dependence of the $\mathrm{BSS}$ on $p$ varies strongly with the noise-to-signal ratio $\eta$. This behavior is consistent with a larger contribution of unresolved variability, and hence with a higher effective noise-to-signal ratio, in rainfall forecasting. We discuss this interpretation further in Section \ref{subsec:disc_comparison}.
\begin{table}%[t]
\caption{AUC and LS  for \textbf{hourly accumulated rainfall} forecasts at lead times $h=1$ and $6\,\mathrm{h}$, and probabilities $p \approx 0.04$ and $0.004$.}
\label{tab:RAIN_AUC_logScore}
\centering
\small
\setlength{\tabcolsep}{6pt}

\begin{tabular}{llllll}
\hline
Model & $p$ & \multicolumn{2}{c}{AUC} & \multicolumn{2}{c}{LS} \\
\cline{3-6}
 &  & $1\,\mathrm{h}$ & $6\,\mathrm{h}$ & $1\,\mathrm{h}$ & $6\,\mathrm{h}$ \\
\hline
\multirow{2}{*}{$\mathcal{M}_1$} 
& 0.04 & 0.971 & 0.934 & 0.075 & 0.104 \\
& 0.004 & 0.954 & 0.872 & 0.025 & 0.031 \\
\multirow{2}{*}{$\mathcal{M}_2$} 
& 0.04 & 0.975 & 0.942 & 0.069 & 0.093 \\
& 0.004 & 0.972 & 0.941 & 0.018 & 0.019 \\
\hline
\end{tabular}
\end{table}  
\begin{figure*}[ht]
	\centering
	\includegraphics[width=0.8\linewidth]{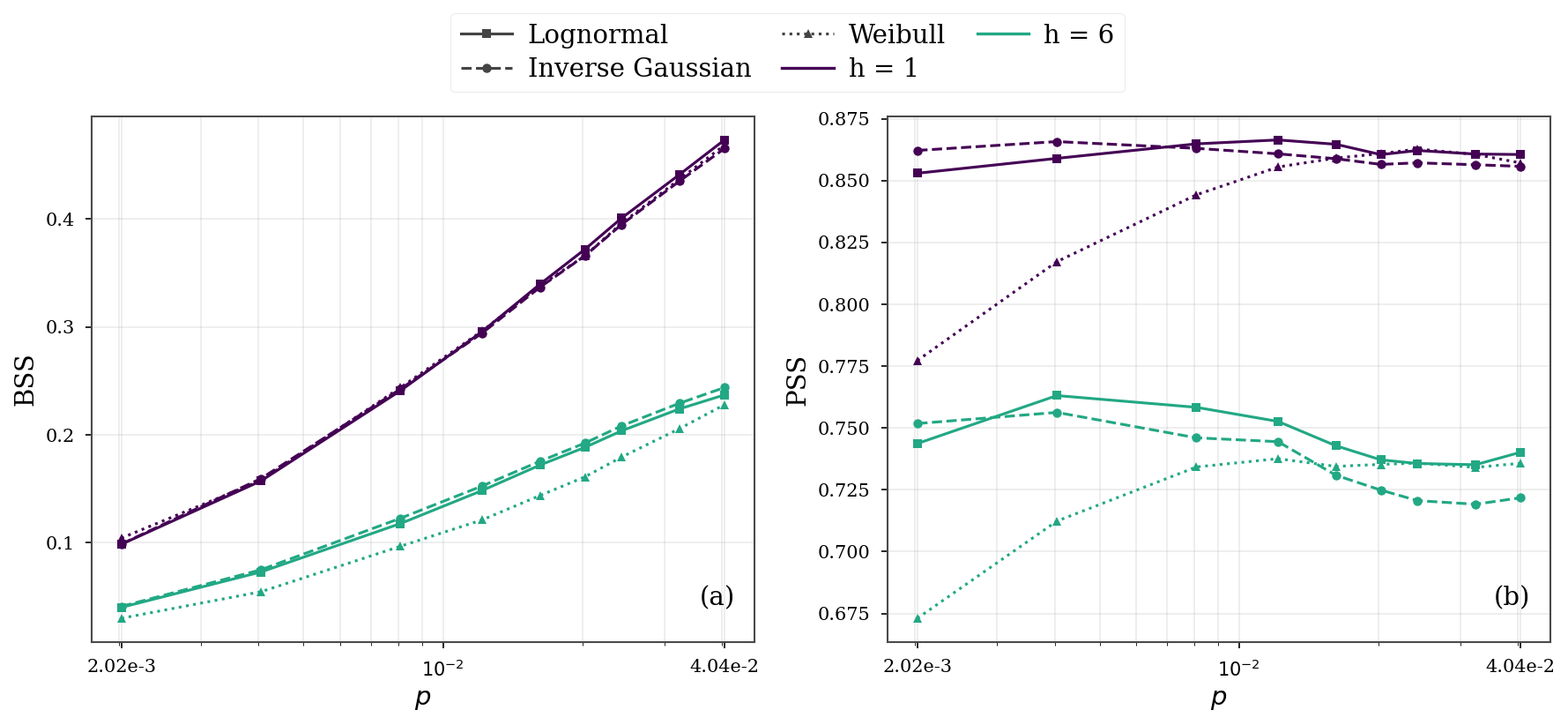}%{Figs/RAIN_results_PSS_BSS_3prob_models_NEW.png}
	\caption{\footnotesize $\text{BSS}$ (panel ~(a)) and $\text{PSS}$ (panel ~(b)) for  \textbf{hourly accumulated rainfall} are shown for different choices of the parametric distribution in model $\mathcal{M}_2$. All the three displayed families are mixed distributions of type ~(\ref{eq:mixture_general}) with three parameters. More specifically, a mixed lognormal ~(\ref{eq:mixture_lognorm}) (symbols (\scalebox{0.7}{$\blacksquare$}) and solid lines), a mixed inverse Gaussian ~(\ref{eq:mixture_ig}) (symbols (\scalebox{0.7}{$\bullet$}) and dashed lines) and a mixed Weibull distribution ~(\ref{eq:mixture_w}) (symbols (\scalebox{0.7}{$\blacktriangle$}) and dotted lines) have been tested. Two different forecast horizons are highlighted: $h=1$ h (violet) and $h=6$ h (green).}
	\label{fig:RAIN_probModels}
    \vspace*{0cm}
\end{figure*}
Table~\ref{tab:RAIN_AUC_logScore} confirms the same model ranking. Model $M_2$ achieves higher AUC and lower LS values than $M_1$ at both probability levels and forecast horizons. The difference is especially clear at $p\simeq0.004$ and $h=6$, where the AUC decreases from $0.941$ for $M_2$ to $0.872$ for $M_1$. These results are consistent with the strong deterioration of the threshold-specific classifier at the rarest rainfall levels. Finally, since one can legitimately question the effect of the distribution choice on the presented results for model ${M}_2$, different mixed distributions have been tested. We compared the mixed lognormal model with mixed inverse-Gaussian and mixed Weibull distributions, all defined in Appendix~\ref{app:mixture_rainfall}. The results are shown in Fig.~\ref{fig:RAIN_probModels} and
indicate that the choice of parametric family does not significantly affect the results, even if the Weibull distribution appears somewhat less suitable.

\subsection{Discussion}

\subsubsection{Comparative performance of wind and accumulated rainfall predictions}\label{subsec:disc_comparison}
For both wind speed and hourly accumulated rainfall, the empirical results are qualitatively consistent with the mechanism isolated by the toy model: $\mathcal M_2$ generally outperforms $\mathcal M_1$, with a relative advantage that tends to increase as the exceedance probability decreases. Rainfall forecasting nevertheless appears more challenging, with systematically lower probabilistic and classification scores. The PSS also does not increase toward rare events for rainfall than for wind speed, indicating a that discrimination does not increase with $p\to0$ as predicted by the idealized location-shift mechanism. This suggests that rainfall predictability may be driven more by the occurrence probability $p_{\mathrm{wet}}(t)$ than by variations in conditional intensity, so that predictors distinguish wet from dry conditions more effectively than moderately heavy from extreme rainfall.

We also note that forecast performance decreases more strongly from $h=1$ to $h=6$ for rainfall than for wind speed, indicating a faster loss of predictable information with forecast horizon. In terms of the toy model, this is qualitatively consistent with a stronger increase in the effective noise-to-signal ratio $\eta$ for rainfall. This behavior may also be related to the validation loss of the rainfall models, which reaches its minimum after only a few epochs and then increases, suggesting that the limited generalizable signal is learned rapidly before further training leads to overfitting.

\subsubsection{Role of the parametric family within the $\mathcal{M}_2$ approach} 
\label{subsec:modelspec} 
Selecting a parametric family for the conditional distribution is challenging because only one outcome is observed for each high-dimensional atmospheric state. Estimation therefore requires assumptions that allow information to be shared across related states, for example through a parametric family or a regularized statistical model. Strictly proper scoring rules provide a principled framework for comparing probabilistic forecasts, since their expected value is optimized by the true predictive distribution \citep{gneiting2007strictly}. However, in finite samples, several candidate families may yield similar average scores over the covariates and thresholds represented in the test data. Under model misspecification, their ranking may also depend on the selected scoring rule and prediction task. Similar scores therefore indicate limited empirical distinguishability rather than structural equivalence between the candidate families. 

In the present rainfall application, Fig.~\ref{fig:RAIN_probModels} shows that the mixed lognormal, inverse-Gaussian and Weibull distributions yield similar BSS (and also PSS) values over the considered threshold range, although the Weibull model performs slightly worse. A similar limited sensitivity to the parametric family was found for wind-speed forecasts in \citet{BaggioMuzy2024}, where M-Rice, Weibull and Gamma predictive distributions produced comparable probabilistic and deterministic scores. These results suggest that, for the datasets and families examined here, a substantial part of the predictive information is associated with changes in bulk properties of the conditional distribution. This interpretation provides a qualitative connection with the Gaussian location model of Section~\ref{sec:theormod}, in which predictable changes are represented exclusively by the latent mean $\mu_t$ defined in Eq.~\eqref{eq:def_mu}. In the applications, the corresponding changes may involve location, dispersion, rainfall-occurrence probability, or other shape parameters. When threshold exceedances are mainly driven by such predictable changes in the bulk of the conditional distribution, several suitable parametric families may produce similar exceedance probabilities despite differences in their detailed shapes. In particular, when the residual conditional variability is small relative to the predictable variation of the distribution, the mechanism represented by the Gaussian toy model may remain qualitatively informative even when the target variable is non-Gaussian. The limited sensitivity observed here therefore supports this interpretation for the present applications, but does not establish it as a universal property of distributional forecasting.
In particular, an inappropriate parametric family may substantially degrade the performance of $M_2$ and remove its advantage over direct classification.

\conclusions \label{sec:conclusion}
In this study, we systematically compared two distinct paradigms for short-term probabilistic forecasting of atmospheric threshold exceedances at a given location: direct binary classification, which frames exceedance as a Bernoulli outcome, and full-distribution modeling, which retains all input data to estimate the conditional probability law of the target variable.  To quantify the statistical consequences of this distinction, we introduced a tractable Gaussian location model inspired by \citet{lerch2017forecaster}. 
Under the regularity and approximation conditions stated in Appendix~\ref{app:asympt_error}, this framework provides explicit expressions for the probability-estimation errors and their effects on the BSS, LS, and PSS as the marginal event probability $p$ decreases. 

Within this idealized setting, thresholding the continuous response reduces the information available for estimating rare-event probabilities. Consequently, as $p\to0$, the estimation error of the full-distribution approach becomes asymptotically smaller than that of direct classification. This difference is reflected in all three verification measures considered. In the model, increasingly rare exceedances become progressively associated with large excursions of the predictable location component, so that their intrinsic discrimination increases. The full-distribution model can exploit this increasing discrimination, whereas direct classification eventually deteriorates because its threshold-specific positive labels become increasingly scarce. The BSS decreases toward rarer events for both approaches, indicating that maintaining probabilistic skill relative to climatology becomes more difficult, but this deterioration is stronger for direct classification. The logarithmic score leads to the same relative ranking. Thus, although the scores have different dependences on $p$, they consistently indicate an increasing relative advantage of the full-distribution approach in the rare-event regime. The noise-to-signal ratio $\eta^2$ defined in Eq.~\eqref{eq:ns_ratio} controls the intrinsic predictability of the events at a given probability level. When $\eta^2$ is small, the predictable location component dominates the residual noise, the conditional probabilities $p_t$ frequently take values close to zero or one and events and non-events are highly discriminable. As $\eta^2$ increases, $p_t$ approaches the climatological probability $p$ and $\operatorname{Var}(p_t)$ decreases. The conditional probability therefore becomes easier to estimate in absolute terms, while carrying less covariate-dependent information for distinguishing events from non-events. Accordingly, both the BSS and PSS decrease as $\eta^2$ increases and vanish in the pure-noise limit, even though the absolute probability-estimation errors may also decrease.

The applications to wind speed and hourly accumulated rainfall in the MeteoNet dataset recover the main qualitative implication of the theoretical analysis: the full-distribution approach outperforms direct classification for all the considered metrics, and its relative advantage becomes more pronounced for rarer thresholds. This result is consistent with the ability of distributional models to learn predictable changes from moderate and non-extreme observations that are discarded when the response is reduced to a binary indicator. The empirical results nevertheless also illustrate the limits of the Gaussian location mechanism. For wind speed, the increase in discrimination toward rarer events is qualitatively similar to that predicted by the toy model. For rainfall, in contrast, $\mathrm{PSS}_2$ remains approximately constant over the probability range considered, whereas $\mathrm{PSS}_1$ decreases. The increasing relative advantage of the distributional approach therefore results primarily from the deterioration of direct classification rather than from an improvement in the discrimination of the distributional model. Increasingly extreme rainfall events thus do not become progressively more discriminable in the same way as in the Gaussian location model. This difference may reflect the importance of other predictable characteristics, including the conditional wet-event probability $p_{\mathrm{wet}}(t)$ and the conditional distribution of positive rainfall amounts. In particular, the available covariates may primarily identify atmospheric conditions conducive to rainfall, through $p_{\mathrm{wet}}(t)$, without providing increasingly distinctive precursors of the most extreme rainfall intensities. Accordingly, the covariate configurations associated with extreme rainfall events may not become progressively more distinct from those associated with less intense positive rainfall events as the threshold increases.

More broadly, the present approach provides an analytical guide for investigating, in future research, how information retention affects rare-event forecasting when predictability is carried by characteristics of the conditional distribution other than its location. Natural extensions include models with a predictable conditional variance, state-dependent shape like tail parameters or, in mixed distributions, the class probability of occurrence, such as $p_{\mathrm{wet}}(t)$ for rainfall. Such extensions would help determine which components of a conditional distribution contribute most strongly to exceedance predictability and how the information loss induced by thresholding depends on the underlying rare-event mechanism. They would also provide a more general framework for interpreting differences between the theoretical and empirical dependence of verification scores on $p$. A related issue is model misspecification. The similar performances obtained here for several reasonably suitable parametric families indicate limited empirical sensitivity to the selected family over the datasets and threshold ranges considered. This observation suggests that a substantial fraction of the relevant predictive information is associated with broad changes in the conditional distribution, but it does not imply robustness to arbitrary misspecification. A family that cannot represent the support, occurrence probability, dispersion, or relevant tail behavior of the target variable may substantially degrade the performance of the full-distribution approach and remove its advantage over direct classification. Future work should therefore extend the analytical framework to heteroskedastic, mixed, state-dependent, and heavy-tailed conditional models and examine how distributional misspecification interacts with the information retained by continuous-response modeling. In particular, integrating Extreme Value Theory into deep distributional forecasting models may help address situations in which exceedances are driven by predictable changes in conditional tail behavior rather than primarily by shifts in the bulk of the distribution.

\vspace*{-0.7cm}
\codedataavailability{
  The meteorological data used in this study originate from the MeteoNet database \citep{larvor2021meteonet}, originally developed by Météo-France. The reference version of the dataset utilized in this work is hosted on the Harvard Dataverse and can be accessed at \url{https://doi.org/10.7910/DVN/NCKRZ2}. The complete Python source code for data preprocessing, model implementation and analysis scripts, along with a minimal self-contained example dataset and an interactive Jupyter Notebook illustrating all aspects of code usage (including data preparation, model training and visualization), is publicly available under the MIT License on Zenodo at \url{https://doi.org/10.5281/zenodo.20327672} \citep{saphir_predict_software}.
}
\vspace*{-1.2cm}
\authorcontribution{Roberta Baggio contributed to code development, experiment design, model execution, result analysis and manuscript writing. Jean-François Muzy contributed to the theoretical analysis, model setup, experiment design, code set up and development and manuscript writing.} %% this section is mandatory
\vspace*{-1.2cm}
\competinginterests{The authors declare that they have no competing interests.} 

\vspace*{-1.2cm}
\financialsupport{Both authors were supported in their research by the ANR research grant SAPHIR (ANR-21-CE04-0014).}

\appendix
\section{Computation of prediction error in asymptotic regime}
\label{app:asympt_error}

Let us estimate the prediction error associated with each prediction method when the number of observations $N$ is large enough and we assume standard asymptotic regularity conditions  \citep[see, e.g.,][]{Vaart_1998}. 

Let ${\boldsymbol{\theta}}$ denote the vector of model parameters 
(e.g., the collection of weights and biases in a neural network).
For a given input $X_t$, the model output is written as $\widehat{z}(X_t;{\boldsymbol{\theta}}) \in \mathbb{R}^d.$
This means that we have $\widehat{z} = \widehat{p}$ for $M_1$ and $\widehat{z} = \widehat{\mu}$ for model $M_2$.
Let $\ell(Y,X;{\boldsymbol{\theta}})$ denote the \emph{per-sample loss function}
(e.g., $[Y-\widehat{\mu}(X;\boldsymbol{\theta})]^2$ for the MSE which is equivalent to the Gaussian negative log-likelihood when the variance is known and fixed) from which the 
empirical loss is defined, over the observation times $\{t_i\}_{i=1 \ldots N}$, as:
$$
\mathcal{L}({\boldsymbol{\theta}})_N  = \frac{1}{N}\sum_{i=1}^N \ell(Y_{t_i},X_{t_i};{\boldsymbol{\theta}}) \;.
$$
Under correct specification, we assume that there exists a unique true parameter value, 
$\boldsymbol{\theta}_0 = \arg\min_{\boldsymbol{\theta}} \mathbb E\left[ \mathcal L_N(\boldsymbol{\theta}) \right]$, 
such that $\widehat z(X_t;\boldsymbol{\theta}_0)=z_0(X_t)$, where $z_0(X_t)$ is the data-generating function, namely $\mu(X_t)$ for $M_2$ and $P(X_t)$ for $M_1$. Although correct specification may be unrealistic in some applications, it provides a useful regular framework for comparing $M_1$ and $M_2$.
We further assume that the estimator $\widehat{\boldsymbol{\theta}}$ minimizing $\mathcal L_N(\boldsymbol{\theta})$ is consistent and satisfies the usual asymptotic normality property
\begin{equation}
\label{eq:asympt_norm}
  \sqrt{N}\,\big(\widehat{{\boldsymbol{\theta}}} - {\boldsymbol{\theta}}_0\big)
  \;\xrightarrow{d}\;
  \mathcal{N}\!\big(0,\, I_{{\boldsymbol{\theta}}}({\boldsymbol{\theta}}_0)^{-1}\big),
\end{equation}
where
$I_{\boldsymbol{\theta}}(\boldsymbol{\theta}_0) = \mathbb E \left[ -\left. \nabla_{\boldsymbol{\theta}}^2 \ell(Y,X;\boldsymbol{\theta}) \right|_{\boldsymbol{\theta}=\boldsymbol{\theta}_0} \right]$ is the Fisher information matrix. Applying the Delta method to the smooth mapping $\widehat{z}(X;{\boldsymbol{\theta}})$ then yields, for each fixed input $X$,
\begin{equation}
\label{eq:asympt_norm_output}
  \sqrt{N}\,\big(\widehat{z}(X) - z_0(X)\big)
  \;\xrightarrow{d}\;
  \mathcal{N}\!\Big(
    0,\,
    G(X)^\top \, I_{\boldsymbol{\theta}}({\boldsymbol{\theta}}_0)^{-1}\, G(X)
  \Big),
\end{equation}
where $G(X) = \left. \nabla_{\boldsymbol{\theta}} \widehat z(X;\boldsymbol{\theta}) \right|_{\boldsymbol{\theta}=\boldsymbol{\theta}_0}$
is the Jacobian of the model output with respect to the parameters,
evaluated at ${\boldsymbol{\theta}}_0$.
Under these conditions, we can estimate the error associated with each method.

Throughout this appendix, the symbol $\approx$ denotes a first-order delta-method approximation whereas 
$\simeq$ denotes a modelling approximation used at finite $p$, 
such as the factorization of the Fisher information or the replacement of a local variance function by a constant as 
considered below. Finally, $f(p)\sim g(p)$ means $f(p)/g(p)\to1$ as $p\to0$.
We also need to define the function $\Phi_k(x_1,\ldots,x_k;r)$ as the CDF of a $k$-dimensional standard Gaussian vector with common pairwise correlation $r$, evaluated at $(x_1,\ldots,x_k)$. When all arguments are equal, we use the shorthand 
\begin{equation} 
\Phi_k(x;r) := \Phi_k(x,\ldots,x;r). 
\label{eq:Phi_k_diagonal} 
\end{equation}
Let us also mention that the large-sample approximations are first derived for fixed $p>0$ as $N\to\infty$. The rare-event behavior of the resulting expressions is subsequently examined as $p\to0$. If $p=p_N$ decreases jointly with $N$, additional regularity conditions are required, as specified below for $M_1$.

\subsection{Model $M_2$ in class $\mathcal{M}_2$}

For model $M_2\in\mathcal{M}_2$, we write
$\widehat z(X_t;\boldsymbol{\theta})
=\widehat\mu(X_t)=M_2(X_t;\boldsymbol{\theta})$, where $M_2$ is defined
in Eq.~\eqref{eq:param_estimation}. In the Gaussian model considered
here, the conditional mean is the only parameter depending on the
covariates, while $\sigma^2$ is known and constant. The per-observation
Fisher information and the resulting asymptotic variance are
\begin{equation}
	\label{eq:M2_information}
	\begin{aligned}
		I_2(\boldsymbol{\theta}_0) &= \sigma^{-2}J_2(\boldsymbol{\theta}_0), \\
		J_2(\boldsymbol{\theta}_0) &= \mathbb{E}_{X_t} \left[ g_2(X_t)g_2(X_t)^\top \right], \\
		\operatorname{Var}_{\widehat{\boldsymbol{\theta}}} \left( \widehat\mu(X_t)\mid X_t \right) &= \frac{\sigma^2}{N}V_2(X_t) + o(N^{-1}), \\
		V_2(X_t) &= g_2(X_t)^\top J_2(\boldsymbol{\theta}_0)^{-1} g_2(X_t).
	\end{aligned}
\end{equation}
where
$g_2(X_t)=
\nabla_{\boldsymbol{\theta}}
M_2(X_t;\boldsymbol{\theta}_0)$.

Applying the delta method to the exceedance probability defined
previously gives
\begin{equation}
	\label{eq:e2_1}
	\begin{aligned}
		& \operatorname{Var}_{\widehat{\boldsymbol{\theta}}} \left( \widehat p^{(2)}(X_t)\mid X_t \right) 
		= \frac{V_2(X_t)}{N} \phi^2\left( \frac{Q_p-\mu(X_t)}{\sigma} \right) + o(N^{-1}), \\
		& \mathcal E_2 
		= \frac{1}{N} \mathbb E_{X_t} \left[ V_2(X_t) \phi^2\left( \frac{Q_p-\mu(X_t)}{\sigma} \right) \right] + o(N^{-1}).
	\end{aligned}
\end{equation}
Let ${\overline V}_2(u) =
\mathbb E[V_2(X_t)\mid\mu(X_t)=u]$.
Since $\mu(X_t)$ is Gaussian with zero mean and variance $s^2$,
Eq.~\eqref{eq:e2_1} becomes
\begin{equation}
\label{eq:error_p2}
\mathcal E_2
\approx
\frac{1}{N(2\pi)^{3/2}s}
\int_{\mathbb R}
{\overline V}_2(u)
\exp\left(-\frac{u^2}{2s^2}\right)
\exp\left(-\frac{(Q_p-u)^2}{\sigma^2}\right)
\,du.
\end{equation}

Equation~\eqref{eq:error_p2} can be evaluated analytically for several
forms of ${\overline V}_2(u)$, including polynomial functions. Such
parameterizations, however, introduce additional coefficients that are
not central to the rare-event scaling studied here. We therefore adopt
the more parsimonious approximation
\begin{equation}
\label{eq:constant_V2}
\overline{V}_2(u)\simeq V_2,
\end{equation}
where $V_2>0$ is constant. 
To assess this approximation, we performed additional numerical experiments in which ${\overline V}_2(u)$ was estimated from the conditional prediction variance using Eq.~\eqref{eq:e2_1}: $ V_2(u) \approx \frac{ N\,\operatorname{Var}_{\widehat{\boldsymbol{\theta}}} \left( \widehat p^{(2)}(X_t) \mid \mu(X_t)=u \right) }{ \phi^2\left((Q_p-u)/\sigma\right) }$.
The estimated function varied only weakly with $u$ over the range contributing materially to the prediction error, 
supporting its approximation by a constant in the present experiments.
Using the expressions of $Q_p$ and $\eta$ defined previously,
Eq.~\eqref{eq:error_p2} then gives
\begin{equation}
\label{eq:error_p2_A}
\mathcal E_2
\simeq
\frac{V_2\eta}
{2\pi N\sqrt{2+\eta^2}}
\exp\left[
-\frac{1+\eta^2}{2+\eta^2}
\left\{\Phi^{-1}(p)\right\}^2
\right].
\end{equation}
Since
$\exp[-\{\Phi^{-1}(p)\}^2]
\sim4\pi p^2\ln(1/p)$,
it follows that
\begin{equation}
	\label{eq:error_p2_final}
	\begin{aligned}
		\mathcal E_2 &\sim \frac{K_2(\eta)}{N} \, p^{\frac{2+2\eta^2}{2+\eta^2}} \left[\ln(1/p)\right]^{\frac{1+\eta^2}{2+\eta^2}}, \\
		\text{with } K_2(\eta) &= \frac{V_2\eta}{2\pi\sqrt{2+\eta^2}} \, (4\pi)^{\frac{1+\eta^2}{2+\eta^2}}.
	\end{aligned}
\end{equation}
Thus, up to logarithmic corrections,
$\mathcal E_2$ scales as $\eta p/N$ for $\eta\ll1$ and as $p^2/N$
for $\eta\gg1$.

\subsection{Model $M_1$ in class $\mathcal{M}_1$}

For model $M_1\in\mathcal{M}_1$, the output is the direct estimate
$\widehat p^{(1)}(X_t;\boldsymbol{\theta})$ and the loss is the binary
cross-entropy given in Eq.~\eqref{eq:bce_loss}. 
Writing the sigmoid function, $\operatorname{Sig} (u) = \frac{1}{1+e^{-u}}$, one has
$\widehat p^{(1)}(X_t;\boldsymbol{\theta}) =
\operatorname{Sig} \left[
L_p(X_t;\boldsymbol{\theta})\right]$,
the subscript $p$ emphasizes that a separate classifier is fitted for
each threshold. At the population parameter
$\boldsymbol{\theta}_{0,p}$, correct specification implies
$\widehat p^{(1)}(X_t;\boldsymbol{\theta}_{0,p})=p_t$.
Using $\operatorname{Sig}'(u)
=\operatorname{Sig}(u)\{1-\operatorname{Sig}(u)\}$,
the probability gradient and the binary cross-entropy score are
\begin{equation}
	\label{eq:M1_gradients}
	\begin{aligned}
		&\left. \nabla_{\boldsymbol{\theta}} \widehat p^{(1)}(X_t;\boldsymbol{\theta}) \right|_{\boldsymbol{\theta}=\boldsymbol{\theta}_{0,p}} 
		= p_t(1-p_t) g_{1,p}(X_t), \\
		& \nabla_{\boldsymbol{\theta}}\ell 
		= \left( \widehat p_t^{(1)} - I_{t+h} \right) \nabla_{\boldsymbol{\theta}}L_p.
	\end{aligned}
\end{equation}
where
$g_{1,p}(X_t)=
\nabla_{\boldsymbol{\theta}}
L_p(X_t;\boldsymbol{\theta}_{0,p})$.
The corresponding Fisher information is therefore
\begin{equation}
\label{eq:I_M1}
I_{1,p}(\boldsymbol{\theta}_{0,p})
=
\mathbb E_{X_t}
\left[
p_t(1-p_t)
g_{1,p}(X_t)g_{1,p}(X_t)^\top
\right].
\end{equation}
Combining Eq.~\eqref{eq:M1_gradients} with
Eq.~\eqref{eq:asympt_norm_output} gives
\begin{equation}
\label{eq:e1_1}
\mathcal E_1
\approx
\frac{1}{N}
\mathbb E_{X_t}
\left[
p_t^2(1-p_t)^2
g_{1,p}(X_t)^\top
I_{1,p}(\boldsymbol{\theta}_{0,p})^{-1}
g_{1,p}(X_t)
\right].
\end{equation}
Equation~\eqref{eq:e1_1} involves the full inverse Fisher information matrix, including its off-diagonal entries and therefore does not assume uncorrelated parameter estimates. Without further approximation, the Fisher matrix and the resulting quadratic-form expectation can be estimated numerically for each threshold. To obtain a closed-form approximation, we introduce a specific
factorization of the Fisher information by neglecting the matrix-valued
covariance between the conditional Bernoulli variance factor
$p_t(1-p_t)$ and the local sensitivity matrix
$g_{1,p}(X_t)g_{1,p}(X_t)^\top$. This gives
\begin{equation}
	\label{eq:I_M1_factorized}
	\begin{aligned}
		& I_{1,p}(\boldsymbol{\theta}_{0,p}) \simeq C_p J_{1,p}, \; \mathrm{with} \\
		& C_p = \mathbb{E}_{X_t}[p_t(1-p_t)] = p - \Phi_2(z_p; r), \\
		& J_{1,p} = \mathbb{E}_{X_t} \left[ g_{1,p}(X_t)g_{1,p}(X_t)^\top \right].
	\end{aligned}
\end{equation}
and where $z_p=\Phi^{-1}(p)$ and $r=(1+\eta^2)^{-1}$. The expression of $C_p$ follows from $\mathbb E[p_t]=p$ and $\mathbb E[p_t^2]=\Phi_2(z_p;r)$.
This factorization simplifies the dependence on the conditional Bernoulli variance factor but the model-dependent matrix $J_{1,p}$ would still generally require numerical evaluation. Defining
\begin{equation}
V_{1,p}(X_t)
=
g_{1,p}(X_t)^\top
J_{1,p}^{-1}
g_{1,p}(X_t),
\end{equation}
Eqs.~\eqref{eq:e1_1} and \eqref{eq:I_M1_factorized} yield:
\begin{equation}
	\label{eq:e3_1_general}
	\begin{aligned}
		&\operatorname{Var}_{\widehat{\boldsymbol{\theta}}} \left( \widehat p^{(1)}(X_t)\mid X_t \right) 
		\simeq \frac{p_t^2 (1-p_t)^2 V_{1,p}(X_t)}{N C_p}, \\
		&\mathcal E_1 
		\simeq \frac{1}{NC_p} \mathbb E_{X_t} \left[ p_t^2(1-p_t)^2 V_{1,p}(X_t) \right].
	\end{aligned}
\end{equation}
It is noteworthy that in a joint limit $p=p_N\to0$, regular asymptotic normality for $M_1$ additionally requires, for fixed $r<1$, 
\begin{equation} \label{eq:cond} 
NC_{p_N} = N\left[ p_N-\Phi_2(z_{p_N};r) \right] \sim Np_N \longrightarrow\infty. 
\end{equation} 
Thus, regimes with a bounded or vanishing expected number of exceedances are not covered by the regular Gaussian asymptotics.

As for $M_2$, Eq.~\eqref{eq:e3_1_general} can be evaluated 
for more general local variance functions, including functions of the latent signal. 
To avoid additional fitted coefficients, we adopt the parsimonious approximation 
\begin{equation} 
\label{eq:constant_V1} 
V_{1,p}(X_t)\simeq V_1, 
\end{equation} 
where $V_1>0$ is taken to be independent of both $X_t$ and $p$. 
Additional numerical experiments based on Eq.~\eqref{eq:e3_1_general} showed only a weak dependence 
on $u$ of the estimated conditional average $\overline V_{1,p}(u) = \mathbb E[V_{1,p}(X_t)\mid\mu(X_t)=u]$ over the relevant range, supporting the constant approximation used here.

Since
$\mathbb E[p_t^k]=\Phi_k(z_p;r)$ as defined in Eq.~(\ref{eq:Phi_k_diagonal}) and denoting $\Delta(z_p;r) = \Phi_2(z_p;r)-2\Phi_3(z_p;r)+\Phi_4(z_p;r)$ one obtains
\begin{equation}
\label{eq:error_p1_finite}
\mathcal E_1
\simeq
\frac{V_1}{N}
\frac{\Delta(z_p;r)}
{p-\Phi_2(z_p;r)}
\simeq 
\frac{V_1}{N(p-\Phi_2(z_p;r))}\Phi_2(z_p;r),
\end{equation}
where the last term uses the fact that, for small $p$, $\Delta(z_p;r)\sim\Phi_2(z_p;r)$. 
As $p\to0$, since
\begin{equation}
\label{phi2_asympt}
\Phi_2(z_p;r)
\sim
C(r)
p^{\frac{2}{1+r}}
\left[\ln(1/p)\right]^{-\frac{r}{1+r}},
\end{equation}
with $r=(1+\eta^2)^{-1}$ and $C(r) = \frac{(1+r)^2 (4\pi)^{-\frac{r}{1+r}}}{\sqrt{1-r^2}}$
we have the following asymptotic behavior:
\begin{equation}
\label{eq:error_p1_final}
\mathcal E_1
\sim
\frac{K_1(\eta)}{N}
p^{\frac{\eta^2}{2+\eta^2}}
\left[\ln(1/p)\right]^{
-\frac{1}{2+\eta^2}},
\;
K_1(\eta)
=
V_1
C\left(
\frac{1}{1+\eta^2}
\right).
\end{equation}
Thus, up to logarithmic corrections,
$\mathcal E_1$ scales as $N^{-1}$ for $\eta\ll1$ and as $p/N$ for
$\eta\gg1$.

Finally, Eqs.~\eqref{eq:error_p1_final} and
\eqref{eq:error_p2_final} give
\begin{equation}
\label{eq:error_ratio_final}
\frac{\mathcal E_2}{\mathcal E_1}
\sim
\frac{K_2(\eta)}{K_1(\eta)}
p\ln(1/p).
\end{equation}
Hence, within the regular asymptotic framework and up to logarithmic
corrections, the distributional approach gains one leading power of
$p$ over direct classification.

\section{Relative BS and LS performances of $\mathcal{M}_1$ and $\mathcal{M}_2$}
\label{app:logscore}

We can exploit the conditional variance formulas established in Appendix~\ref{app:asympt_error} to derive the asymptotic behavior BS and LS scores associated with predictions of models $M_1$ and $M_2$. As before, we pay particular attention to  the rare-event regime ($p \to 0$).
We denote ${\widehat{p}}_t^{(k)}$ denote the estimator of the exceedance probability $p_t = \mathrm{Pr}(Y_{t+h} > Q_p \mid X_t)$ with model $M_k$ where $k=1$ or $2$.
According to Eq.~(\ref{eq:def_Brier}), the BS of $M_k$, reads (by replacing the average over observations by the mathematical expectation over the joint law of $(\widehat{{\boldsymbol{\theta}}},X_t,\nu_{t+h})$
or equivalently of  $(\widehat{{\boldsymbol{\theta}}},p_t,I_{t+h})$
\begin{align*}
\mathrm{BS}_k
&=
\mathbb E\left[
\left(\widehat p_t^{(k)}-I_{t+h}\right)^2
\right]
\\
&=
\mathbb E\left[
\left(\widehat p_t^{(k)}-p_t\right)^2
\right]
+
\mathbb E\left[
\left(p_t-I_{t+h}\right)^2
\right]
\\
&\quad+
2\mathbb E\left[
\left(\widehat p_t^{(k)}-p_t\right)
\left(p_t-I_{t+h}\right)
\right]
\\
&=
\mathcal E_k+
\mathbb E_{p_t}\left[p_t(1-p_t)\right].
\end{align*}
Indeed, provided that the verification observation is independent of
the training sample, the cross term vanishes because $ \mathbb E\left[
p_t-I_{t+h}\mid p_t,\widehat{\boldsymbol{\theta}} \right]=0$. 
Moreover, since $p_t=\Pr(I_{t+h}=1\mid X_t)$, the tower property gives $\mathbb E[I_{t+h}\mid p_t]=p_t$. Together with $I_{t+h}^2=I_{t+h}$, this yields $\mathbb E[(p_t-I_{t+h})^2\mid p_t]=p_t(1-p_t)$.
Thanks to Eqs.~(\ref{E1}) and ~(\ref{E2}) we finally have the exact relationship between the BS of $M_k$ and the previously computed asymptotic error $\mathcal{E}_k$ that reads:
\begin{equation}
\label{BS_k_vs_E_k}
      BS_k = \mathcal{E}_k + p - \Phi_2\left(\Phi^{-1}(p),r\right)
\end{equation}
with $r = \frac{1}{1+\eta^2}$.
It results from ~(\ref{eq:def_BSS}), that the Brier Skill of each model $M_k$ is :
\begin{equation}
\label{BSS_k_vs_E_k}
      BSS_k = 1-\frac{\mathcal{E}_k + p - \Phi_2\left(\Phi^{-1}(p),r\right)}{p(1-p)}
\end{equation}
Using the asymptotic behavior of $\Phi_2$ in Eq.~\eqref{phi2_asympt}, we obtain 
\begin{equation} 
\label{eq:BSS_k_asympt2} 
BSS_k \simeq C(r) p^{\frac{\eta^2}{2+\eta^2}} \bigl[\ln(1/p)\bigr]^{-\frac{1}{2+\eta^2}} - \frac{\mathcal E_k}{p}, \end{equation} 
up to factors tending to one as $p\to0$.

The logarithmic score corresponds to the binary cross-entropy defined in Eq.~(\ref{eq:bce_loss}) which expectation gives
\begin{equation*}\mathrm{LS}_k  =  - {\mathbb{E}}_{\widehat{{\boldsymbol{\theta}}}}{\mathbb{E}}_{X_t} {\mathbb{E}}_{I_{t+h} \mid X_t} \left[  I_{t+h} \ln\left( {\widehat{p}}_t^{(k)} \right) + \left(1 - I_{t+h} \right) \ln \left( 1 - {\widehat{p}}_{t}^{(k)}  \right) \right]
\end{equation*} which leads to:
\begin{equation*}
     \mathrm{LS}_k  = - {\mathbb{E}}_{\widehat{{\boldsymbol{\theta}}}} {\mathbb{E}}_{X_t} \left[  p_t \ln\left( {\widehat{p}}_t^{(k)} \right) + \left(1 - p_t \right) \ln \left( 1 - {\widehat{p}}_{t}^{(k)}  \right) \right] 
\end{equation*}
where, as before, ${\widehat{p}}^{(k)}_t =  {\widehat{p}}^{(k)}( X_t ; \widehat{{\boldsymbol{\theta}}})$ and $I_{t+h} = I_{t+h}(Q_p)$ is defined in Eq.~(\ref{eq:def_Iq}).
In this context, in order to compare methods $M_1$ and $M_2$, one can  
evaluate 
$$\Delta \mathrm{LS}_{1,2} = \mathrm{LS}_1-\mathrm{LS}_2
$$
where a positive value means that $M_2$ performs better than $M_1$.
$\Delta \mathrm{LS}_{1,2}$ can be conveniently expressed as:
$\Delta \mathrm{LS}_{1,2} = {\mathcal R}_1-{\mathcal R}_2$
where ``excess risk'' ${\mathcal R}_k =  D_{KL} \Big(p_t \, \| \, {\widehat{p}}_t^{(k)} \Big) $ represents the Kullback-Leibler divergence 
with respect to the true probability, namely: 
\begin{equation}
\label{eq:def_Rk}
 {\mathcal R}_k =  {\mathbb{E}}_{\widehat{{\boldsymbol{\theta}}}} {\mathbb{E}}_{X_t} \left[
p_t\ln \frac{p_t}{{\widehat{p}}_t^{(k)}} +
(1-p_t)\ln \frac{1-p_t}{1-{\widehat{p}}_t^{(k)}}
\right] \; .
\end{equation}
Under consistency, $\widehat p_t^{(k)}$ converges in probability to $p_t$ as $N\to\infty$. A second-order Taylor expansion of the KL divergence around $p_t$ then gives, under the negligible-bias approximation, 
\begin{equation} 
\mathcal R_k \approx \frac{1}{2} \mathbb E_{X_t} \left[ \frac{ \operatorname{Var}_{\widehat{\boldsymbol{\theta}}} \left( \widehat p_t^{(k)}\mid X_t \right) }{ p_t(1-p_t) } \right]. 
\end{equation}
The conditional variance expressions in Eqs.~\eqref{eq:e3_1_general} and~\eqref{eq:e2_1} therefore yield:
\begin{align}
    \mathcal{R}_1 &\approx \frac{1}{2NC_p} \mathbb{E}_{X_t} \left( p_t(1-p_t) V_1(X_t) \right) \nonumber \\
    \mathcal{R}_2 &\approx \frac{1}{2N} \mathbb{E}_{X_t} \left( \frac{V_2(X_t) \phi^2 \left( \frac{Q_p-\mu(X_t)}{\sigma}\right)}{\Phi \left(\frac{\mu(X_t)-Q_p}{\sigma} \right) \left[1-\Phi \left(\frac{\mu(X_t)-Q_p}{\sigma} \right) \right]} \right) \nonumber
\end{align}
where we used the fact that $p_t = \Phi \left(\frac{\mu(X_t)-Q_p}{\sigma} \right)$.
Under the constant-local-variance approximation introduced in Appendix~\ref{app:asympt_error} and using $1-p_t\simeq1$ in the rare-event regime, we obtain:
\begin{align} 
 \mathcal R_1 &\simeq \frac{V_1}{2N}, \\ 
 \mathcal R_2 &\simeq \frac{V_2}{2N} \mathbb E_{X_t} \left[ \frac{ \phi^2\left((Q_p-\mu(X_t))/\sigma\right) }{ \Phi\left((\mu(X_t)-Q_p)/\sigma\right) } \right]. 
\end{align}
where we used the definition of $C_p$, namely $C_p = {\mathbb{E}}(p_t(1-p_t))$. ${\mathcal R}_2$ can be rewritten as ${\mathcal R}_2 = \frac{V_{2}}{2N} \mathbb{E}_\mu [\phi^2(z)/\Phi(z)]$ with $z=(\mu-Q_p)/\sigma$. 
In the region contributing dominantly to the integral as $p\to0$, the argument $z=(\mu-Q_p)/\sigma$ is large and negative, so that Mills' ratio can be used. We thus have $\Phi(z) \sim \phi(z)/|z|$, the integrand simplifies to a linear-Gaussian form $|z|\phi(z)$.
As before, the resulting integral is evaluated using the saddle-point method, dominated by the contribution at $\mu_* = Q_p s^2/(s^2+\sigma^2)$. This leads to a scaling proportional to $p [\Phi^{-1}(p)]^2$. By incorporating the refined asymptotic expansion of the quantile function and setting $\eta^2 = \frac{\sigma^2}{s^2}$, we obtain $\mathcal R_2 \simeq \frac{V_2\eta^2}{(1+\eta^2)N} p\ln\left(\frac{1}{p}\right)$. Combining the two excess risks gives 
\begin{equation} 
\label{eq:delta_LS_asympt} 
\Delta\mathrm{LS}_{1,2} \simeq \frac{1}{2N} \left[ V_1 - \frac{2V_2\eta^2}{1+\eta^2} p\ln\left(\frac{1}{p}\right) \right]. 
\end{equation} 
Therefore, for sufficiently small $p$, $\Delta\mathrm{LS}_{1,2}>0$, showing that $M_2$ outperforms $M_1$ according to the logarithmic score.

\section{Asymptotic Analysis of the Peirce Skill Score (PSS)}
\label{app:pss}

Let us perform the same kind of analysis for PSS score. We can remark, from the definition of ~(\ref{eq:PSS}), the averaged $\mathrm{PSS}_k$ for method $\mathcal{M}_k$ can be written as:
\begin{equation}
\begin{aligned}
\label{eq:pss_k}
	\text{PSS}_k & = {\mathbb{E}}_{\widehat{{\boldsymbol{\theta}}}} \left[  \mathrm{Pr} \left({\widehat{I}}_{t+h}^{(k)} =1 \; \mid \;  I_{t+h}=1 \right)  \right. \\ &  \left. - \mathrm{Pr} \left({\widehat{I}}_{t+h}^{(k)} =1  \; \mid \; I_{t+h}=0 \right) \right]
\end{aligned}
\end{equation}
where $k=1,2$ and ${\widehat{I}}_{t+h}^{(k)}$ is defined as in Eq. ~(\ref{eq:det_pred}), namely,
${\widehat{I}}_{t+h}^{(k)} = 1 \; \mathrm{if} \; {\widehat{p}}_{t}^{(k)} > p \; \mathrm{and} \;  {\widehat{I}}_{t}^{(k)} = 0  \; \mathrm{otherwise}$ with ${\widehat{p}}_t^{(k)} = {\widehat{p}}^{(k)}(X_t;\widehat{{\boldsymbol{\theta}}})$. 
It results that:
\begin{equation*}
\begin{aligned}
&  \mathrm{Pr} \left({\widehat{I}}_{t+h}^{(k)} =1 \; \mid \;  I_{t+h}=1 \right) = \\
&  \int dx \; f_X(x \mid I_{t+h} = 1) \; \mathbb{I}_{\{{\widehat{p}}^{(k)}(x;\widehat{{\boldsymbol{\theta}}}) > p\}}  =\\
& \int dx \; f_X(x \mid I_{t+h} = 1) \;  \mathbb{I}_{\{P(x)-{\widehat{p}}^{(k)}(x;\widehat{{\boldsymbol{\theta}}}) <  P(x)-p \}}  
\end{aligned}
\end{equation*}
where we denoted, at fixed $t$, $f_X(x)$ the pdf of $X_t$ and $P(x) = \Pr(I_{t+h}=1\mid X_t=x) = \mathbb E[I_{t+h}\mid X_t=x]$ so that $P(X_t)=p_t$, as defined in Eq.~\eqref{eq:true_pt}.
By Bayes rule, we have  
$$
\begin{aligned}
& f_X(x \mid I_{t+h} = 1) = \frac{\mathrm{Pr}(I_{t+h} = 1 \mid \; X_t = x) f_X(x)}{\mathrm{Pr}(I_{t+h}=1)}  \\ &  = \frac{P(x) f_X(x)}{p} 
\end{aligned}
$$ and in the same way we have
$$
f_X(x \mid I_{t+h} = 0) = \frac{(1-P(x)) f_X(x)}{1-p} \; .
$$
This thus entails:
\begin{eqnarray*}
&& \mathrm{Pr} \left({\widehat{I}}_{t+h}^{(k)} =1 \; \mid \;  I_{t+h}=1 \right) =  \\
& &  p^{-1} \int P(x) f_X(x) \mathbb{I}_{\{P(x)-{\widehat{p}}^{(k)}(x;\widehat{{\boldsymbol{\theta}}}) <  P(x)-p \}}  
  =  \\
&& p^{-1} {\mathbb{E}}_\mu \left( \Phi\left(\frac{\mu-Q_p}{\sigma} \right)  \mathbb{I}_{\{P(x)-{\widehat{p}}^{(k)}(x;\widehat{{\boldsymbol{\theta}}}) <  \Phi(\frac{\mu-Q_p}{\sigma}) -p \}}  \right)
\end{eqnarray*}
where we made the change of variable $x \rightarrow \mu(x)$ and used Eq. ~(\ref{eq:true_pt}) for the relationship between $P(x)$
and $\mu(x)$, $\Phi$ standing for the standard normal CDF. 
Finally, taking the expectation as respect to the law of $\widehat{{\boldsymbol{\theta}}}$ that is supposed to be asymptotically normal, we obtain:
\begin{eqnarray*}
&& {\mathbb{E}}_{\widehat{{\boldsymbol{\theta}}}} \left[ \mathrm{Pr} \left({\widehat{I}}_{t+h}^{(k)} =1 \; \mid \;  I_{t+h}=1 \right) \right]  \approx \\
   &  & p^{-1} {\mathbb{E}}_\mu \left[ \Phi\left(\frac{\mu-Q_p}{\sigma} \right)  \Phi \left(\frac{\Phi(\frac{\mu-Q_p}{\sigma}) -p}{\sqrt{ \operatorname{Var}_{\widehat{\boldsymbol{\theta}}} \left( \widehat p^{(k)}(X_t)\mid X_t \right) }} \right) \right]  =  \\
 &&  p^{-1} {\mathbb{E}}_{p_t} \left[ p_t  \Phi \left(\frac{p_t-p}{\sqrt{ \operatorname{Var}_{\widehat{\boldsymbol{\theta}}} \left( \widehat p^{(k)}(X_t)\mid X_t \right) }} \right) \right] 
\end{eqnarray*}
Along the same line one can establish that
\begin{eqnarray*}
& & {\mathbb{E}}_{\widehat{{\boldsymbol{\theta}}}} \left[ \mathrm{Pr} \left({\widehat{I}}_{t+h}^{(k)} =1 \; \mid \;  I_{t+h}=0 \right) \right]  \approx \\
&&  (1-p)^{-1} {\mathbb{E}}_{p_t} \left[ (1-p_t)  \Phi \left(\frac{p_t-p}{\sqrt{ \operatorname{Var}_{\widehat{\boldsymbol{\theta}}} \left( \widehat p^{(k)}(X_t)\mid X_t \right) }} \right) \right] 
\end{eqnarray*}
and therefore, from ~(\ref{eq:pss_k}), we obtain the following expression of $\text{PSS}_k$:
	\begin{equation*}
 \text{PSS}_k  \approx \frac{1}{p(1-p)} {\mathbb{E}}_{p_t} \left[ (p_t-p)  \Phi \left(\frac{p_t-p}{\sqrt{ \operatorname{Var}_{\widehat{\boldsymbol{\theta}}} \left( \widehat p^{(k)}(X_t)\mid X_t \right) }} \right) \right]
\end{equation*}
Using Eqs.~\eqref{eq:e3_1_general} and \eqref{eq:e2_1}, together with the constant-local-variance approximation, $1-p_t \simeq 1$ for $M_1$ and $\phi(\Phi^{-1}(p_t)) \sim p_t\sqrt{2\ln(1/p_t)}$ for $M_2$, leads to:
\begin{eqnarray}
\label{eq:pss1_a}
 &  \! \! \! \!  \! \text{PSS}_1   \simeq  \frac{1}{p(1-p)} {\mathbb{E}}_{p_t} \left[ (p_t-p)  \Phi \left(\frac{\sqrt{p-\Phi_2(z_p;r)}(p_t-p)}{p_t \sqrt{V_1}N^{-\frac{1}{2}}} \right) \right] \\ \label{eq:pss2_a}
 & \! \! \! \! \!   \text{PSS}_2  \simeq \frac{1}{p(1-p)} {\mathbb{E}}_{p_t} \left[ (p_t-p)  \Phi \left(\frac{p_t-p}{p_t \sqrt{2 \ln(p_t^{-1})V_2}N^{-\frac{1}{2}}}\right) \right] \; .
\end{eqnarray}
 When $N$ is very large,  more specifically under the regular joint regime of Eq.~\eqref{eq:cond} for $M_1$ and provided that $\ln(1/p)/N\to0$ for $M_2$, $p^{-1} \sigma^{(k)}_N \to 0$, so we can use the expansion of $\Phi(\frac{x}{\varepsilon})$ when
$\varepsilon \to 0$ :
\[
\Phi\Big(\frac{x}{\varepsilon}\Big) = {\mathcal H}(x) + \frac{\varepsilon^2}{2}\delta'(x) + \mathcal{O}(\varepsilon^4),
\]
where ${\mathcal H}(x)$ is the Heaviside function and $\delta'(x)$ is the derivative of the Dirac distribution. 
It results that we have in this regime:
\begin{eqnarray}
\label{PSS_2_1}
 &\text{PSS}_1   \simeq \frac{1}{p(1-p)} \left( {\mathbb{E}}_{p_t} (p_t-p)^+ -\frac{V_{1}}{2N} p f_{p_t}(p) \right)    \\
 \label{PSS_2_2}
 &  \text{PSS}_2  \simeq \frac{1}{p(1-p)} \left( {\mathbb{E}}_{p_t} (p_t-p)^+ -\frac{V_{2}p^2}{N} \ln(\frac{1}{p}) f_{p_t}(p) \right)  
\end{eqnarray}
To evaluate the behavior of $PSS_k$ for $p \ll 1$, we thus have to analyze the asymptotic limits of both $\mathbb{E}_{p_t}[(p_t-p)^+]$ and $f_{p_t}(p)$. Since $p_t = \Phi((\mu_t - Q_p)/\sigma)$ where $\mu_t \sim \mathcal{N}(0, s^2)$, a simple change of variable leads to the density: 
\begin{equation}
f_{p_t}(y) = \frac{\sigma}{s} \frac{\phi\left(\frac{\sigma\Phi^{-1}(y)+Q_p}{s}\right)}{\phi(\Phi^{-1}(y))} \; .
\end{equation}
Evaluating this density at the boundary $y=p$ using the asymptotic relation $\Phi^{-1}(p)^2 \sim 2\ln(1/p)$ leads a power-law behavior:
\begin{equation}f_{p_t}(p) \propto \exp\left( \gamma \frac{\Phi^{-1}(p)^2}{2} \right) \sim p^{-\gamma} \;,
\end{equation}
where the scaling exponent is bounded by $\gamma = \frac{2\sigma}{\sqrt{s^2+\sigma^2}+\sigma} = \frac{2 \eta}{\sqrt{1+\eta^2}+\eta} \in (0,1)$. The expectation $\mathbb{E}[(p_t - p)^+] = \int_p^1 (y-p)f_{p_t}(y)dy$ can be mapped to the original distribution of $\mu_t \sim \mathcal{N}(0, s^2)$. By determining the positivity threshold of the integrand, $L = \Phi^{-1}(p) (\sigma - \sqrt{s^2 + \sigma^2})$ and  applying the bivariate identity $\int \phi(z) \Phi(az + b) dz = \Phi_2(z, \frac{b}{\sqrt{1+a^2}}; \frac{-a}{\sqrt{1+a^2}})$, the integral evaluates to:
\begin{equation*}
\mathbb{E}[(p_t - p)^+] =  p \Phi\left( \frac{L}{s} \right) - \Phi_2\left( \frac{L}{s}, \Phi^{-1}(p); \frac{-s}{\sqrt{s^2 + \sigma^2}} \right) \; .
\end{equation*}

In the limit $p\to0$, the preceding expression yields, 
\begin{equation}
\frac{\mathbb E_{p_t}[(p_t-p)^+]}{p(1-p)}
\simeq
1-K_\eta p^{\kappa^2},
\end{equation}
with $\kappa = \sqrt{1+\eta^2} - \eta$. Previous expression defines the positive coefficient $K_\eta$, depending on $\eta$ but not on $p$. Substituting this result and the asymptotic behavior of
$f_{p_t}(p)$ into Eqs.~\eqref{PSS_2_1} and \eqref{PSS_2_2}, and
neglecting logarithmic corrections, gives
\begin{align}
\label{eq:pss_1_asympt_appendix}
\mathrm{PSS}_1
&\simeq
1-K_\eta p^{\kappa^2}
-\frac{D_{1,\eta}}{N}p^{-\gamma},
\\
\label{eq:pss_2_asympt_appendix}
\mathrm{PSS}_2
&\simeq
1-K_\eta p^{\kappa^2}
-\frac{D_{2,\eta}}{N}p^{1-\gamma},
\end{align}
where $D_{1,\eta},D_{2,\eta}>0$ are independent of $p$ for fixed
$\eta$ and contain the model-dependent local variance factors $V_1$
and $V_2$, respectively. Consequently, under the regular joint regime specified in
Eq.~\eqref{eq:cond}, the performance ratio satisfies
\begin{equation}
\label{eq:PSS_ratio}
\frac{\mathrm{PSS}_1}{\mathrm{PSS}_2}
\simeq
1-\frac{A_{p,\eta}}{N}p^{-\gamma}
+
\mathcal O\left(
\frac{p^{-2\gamma}}{N^2}
\right),
\;
A_{p,\eta}
=
\frac{D_{1,\eta}-D_{2,\eta}p}
{1-K_\eta p^{\kappa^2}}.
\end{equation}
Since $0<\kappa^2<1$, $A_{p,\eta} = D_{1,\eta} \left[ 1+K_\eta p^{\kappa^2} +o\left(p^{\kappa^2}\right) \right] >0$
for sufficiently small $p$. Hence,
$\mathrm{PSS}_1/\mathrm{PSS}_2<1$, showing that $M_2$ outperforms
$M_1$, with a relative PSS difference of order
$(Np^\gamma)^{-1}$. Since $0<\gamma<1$, $Np\to\infty$ implies
$Np^\gamma\to\infty$, so that this difference remains asymptotically
small. At fixed $p$, both methods converge as $N\to\infty$ to the same
Bayes PSS based on the true conditional probability $p_t$.

\section{A toy model: the weighted harmonic model}
\label{app:model1}

Let $\{X_{t}\}_{t=1}^N$ be a sequence of independent random covariate vectors satisfying 
\begin{equation} 
X_t\in\mathbb R^d, \qquad X_t\stackrel{\mathrm{iid}}{\sim}\mathcal N(0,I_d). 
\end{equation} 
Equivalently, the complete covariate sample may be represented by the matrix ${\mathbf X} = (X_1,\ldots,X_N)^\top \in\mathbb R^{N\times d}$. The model generates a scalar signal $\mu_t$ as a weighted sum of centered harmonic transformations of these factors:
\begin{equation}
\label{eq:def_harm}
\mu_t = \sum_{k=1}^d \left[ w_k^{(c)} \left( \cos(X_{t,k}) - e^{-1/2} \right) + w_k^{(s)} \sin(X_{t,k}) \right].
\end{equation}
Here, the constant $e^{-1/2}$ ensures that the cosine terms have zero mean. The coefficients $w_k^{(c)}$ and $w_k^{(s)}$ are fixed model parameters (frozen randomness). They are initialized by drawing from a standard normal distribution and then rescaled by a global factor $\lambda$ to ensure that the theoretical variance of $\mu_t$ matches the target parameter $s^2$. 
%Specifically, the weights satisfy the normalization constraint:
%\begin{equation*}
%\sum_{k=1}^d \left[ (w_k^{(c)})^2 \text{Var}(\cos X) + (w_k^{(s)})^2 \text{Var}(\sin X) \right] = s^2,
%\end{equation*}
%where $\text{Var}(\cos X) = \frac{1}{2}(1 + e^{-2}) - e^{-1}$ and $\text{Var}(\sin X) = \frac{1}{2}(1 - %e^{-2})$.  
\section{Probability distributions for rainfall and wind speed}	
\subsection{The M-Rice probability distribution}
\label{app:M-Rice}
In \citep{BaMuPo11}, the Rice probability distribution (which corresponds to the norm of a two dimensional random vector which components are 2 independent Gaussian random variables of mean $\mu_1$ and $\mu_2$ and of same variance 
$\sigma^2$) has been extended to ``Multifractal Rice" (M-Rice) distribution that accounts for the situation when, 
as observed in turbulence models, this variance $\sigma^2$ is itself stochastic with a log-normal distribution.  
The M-Rice distribution involves 3 parameters, namely the two Rice parameters coming from from Gaussian law $\nu=\sqrt{\nu_1^2+\nu_2^2}$ and $\sigma^2$ and a supplementary parameter, denoted as $\lambda^2$ associated with the variance of the log-normal law. This parameter is referred to, in the literature on turbulence, as the ``intermittency coefficient'' \citep{frisch1995turbulence}. The M-Rice probability density function (PDF) is then:
\begin{equation*}
%\label{def:M-Rice2}
	 f_{\text{MR}}(y)  =  \frac{1}{\sqrt{2 \pi \lambda^2}}  \int e^{-\frac{\omega^2}{2 \lambda^2}}   \frac{y}{e^{2\omega}\sigma^2} e^{-\frac{y^2+\nu^2}{2 e^{2 \omega} \sigma^2}} I_0(\frac{y\nu} {e^{2 \omega} \sigma^2}) \; d \omega \; .
\end{equation*}
where $I_0(z)$ is the order zero modified Bessel function of the first kind.
%As advocated in \citep{BaggioMuzy2024}, this last formula can be fastly evaluated using the a simple Gauss-Hermite quadrature. The M-Rice cumulative distribution function (CDF) or the mean value function can also be obtained along the same way. For the latter, since for a Rice law of parameter $\nu$ and $\sigma^2$, the mean value is $\mu_{\text{R}} = {\displaystyle \sigma {\sqrt {\frac{\pi}{2}}}\,\,L_{\frac{1}{2}}\left(-\frac{\nu^{2}}{2\sigma ^{2}}\right)}$,where $L_{\frac{1}{2}}$ stands for the order $\frac{1}{2}$ Laguerre polynomial,the mean value of a M-Rice distribution reads:
%\begin{equation}
%	\mu_{\text{MR}} (\nu, \sigma, \lambda^2 ) \simeq  \frac{\sigma}{ \sqrt{2}} \sum_{i=1}^n w_i e^{2 \sqrt{2}\lambda y_i} \, L_{\frac{1}{2}}\left(-\frac{e^{2 \sqrt{2} \lambda y_i} \nu^{2}}{2\sigma ^{2}}\right)  \label{eq:mean_mrice} \; 
%\end{equation}
%with $w_{i}={\frac {2^{n-1}n!{\sqrt {\pi }}}{n^{2}[H_{n-1}(y_{i})]^{2}}}$ for a quadrature order $n$, where $H_n(y)$ denotes the n-th Hermite polynomial and $\left{{y}_i\right}$_{i=1,...,n} its roots. For the purpose of this paper, we chose, $n=11$. 
As advocated in \citet{BaggioMuzy2024}, these integrals can be efficiently evaluated using Gauss-Hermite quadrature. The M-Rice CDF and mean can be obtained in the same way. Denoting by $n$ the quadrature order, by $\left\{y_i\right\}_{i=1,…,n}$ the roots of the $n$-th Hermite polynomial $H_n$, and by $w_{i}={\frac {2^{n-1}n!{\sqrt {\pi }}}{n^{2}[H_{n-1}(y_{i})]^{2}}}$ the corresponding Gauss--Hermite weights, the mean of the M-Rice distribution can be numerically approximated as
\begin{equation}
\mu_{\text{MR}} (\nu, \sigma, \lambda^2 ) \simeq  \frac{\sigma}{ \sqrt{2}} \sum_{i=1}^n w_i e^{2 \sqrt{2}\lambda y_i} \, L_{\frac{1}{2}}\left(-\frac{e^{2 \sqrt{2} \lambda y_i} \nu^{2}}{2\sigma ^{2}}\right)  \label{eq:mean_mrice} \; .\end{equation}
In this work, we use the quadrature order $n=11$.

%%%%%%%%%%%%%%%%%%%%%%%%%%%%%%%%%%%%%%%%%%%%%%%%%%%%%%%%%%%%%%%%%%%%
%%%%%%%%%%%%%%%%%%%%%%%%%%%%%%%%%%%%%%%%%%%%%%%%%%%%%%%%%%%%%%%%%%%%

\subsection{Mixed distributions for rainfall}
\label{app:mixture_rainfall}

Mixed distributions combining a point mass at zero with a continuous
positive component are commonly used to model rainfall
\citep{kedem1990estimation}. We define the general mixture density
\begin{equation}
\label{eq:mixture_general}
f(x)
=
(1-p_{\mathrm{wet}})\delta_0(x)
+
p_{\mathrm{wet}}f_+(x;\Pi)\mathbf{1}_{\{x>0\}},
\end{equation}
where $\delta_0$ is the Dirac mass at zero and $f_+(x;\Pi)$ is the
conditional density of positive rainfall. In this context we 
have simple $\Pr(X=0)=1-p_{\mathrm{wet}}$ or equivalently 
$\Pr(X>0)=p_{\mathrm{wet}}$. 
We consider three
two-parameter families for $f_+$, so that the complete mixture model
has three parameters, including $p_{\mathrm{wet}}$.

The first choice is the lognormal density
\begin{equation}
\label{eq:mixture_lognorm}
f_+(x;\mu,\sigma)
=
\frac{1}{x\sigma\sqrt{2\pi}}
\exp\left[
-\frac{(\ln x-\mu)^2}{2\sigma^2}
\right],
\qquad x>0,
\end{equation}
where $\mu$ and $\sigma^2$ control location and dispersion on the
logarithmic scale. The second is the inverse-Gaussian density
\begin{equation}
\label{eq:mixture_ig}
f_+(x;\mu,\lambda)
=
\left(
\frac{\lambda}{2\pi x^3}
\right)^{1/2}
\exp\left[
-\frac{\lambda(x-\mu)^2}{2\mu^2x}
\right],
\qquad x>0,
\end{equation}
where $\mu>0$ is the conditional mean and $\lambda>0$ is a shape
parameter controlling dispersion. The third is the Weibull density
\begin{equation}
\label{eq:mixture_w}
f_+(x;k,\lambda)
=
\frac{k}{\lambda}
\left(
\frac{x}{\lambda}
\right)^{k-1}
\exp\left[
-\left(
\frac{x}{\lambda}
\right)^k
\right],
\qquad x>0,
\end{equation}
where $\lambda>0$ is a scale parameter and $k>0$ controls the shape
and tail behavior. These families are commonly used for rainfall
modeling
(see e.g., \cite{cho2004comparison,Sukrutha2018,wilks1989rainfall,
olivera2019increases,marra2023non}).

%%%%%%%%%%%%%%%%%%%%%%%%%%%%%%%%%%%%%%%%%%%%%%%%%%%%%%%%%%%%%%%%%%%%
\section{CSI and HSS metrics for wind speed and accumulated rainfall}
\label{app:csi_hss}
The Heidke Skill Score (HSS) and the Critical Success Index (CSI) are two additional metrics commonly used in the verification of rare meteorological events \citep{wilks2011statistical}. The HSS measures categorical skill relative to a chance-reference forecast and depends on both observed and forecast marginal frequencies \citep{Jolliffe2004}:
\begin{equation}
\label{eq:HSS}
\text{HSS} = \frac{2(\mathrm{TP}\cdot\mathrm{TN}-\mathrm{FP}\cdot\mathrm{FN})}
{(\mathrm{TP}+\mathrm{FN})(\mathrm{FN}+\mathrm{TN})+(\mathrm{TP}+\mathrm{FP})(\mathrm{TN}+\mathrm{FP})}.
\end{equation}
The CSI focuses on event prediction by ignoring correct non-events:
\begin{equation}
\label{eq:CSI}
\text{CSI} = \frac{\mathrm{TP}}{\mathrm{TP}+\mathrm{FP}+\mathrm{FN}}.
\end{equation}
The CSI is therefore useful when the large number of true negatives associated with rare events would otherwise dominate the verification, but it depends intrinsically on the event base rate. The HSS corrects for chance agreement, although its normalization also depends on the forecast marginal frequency, making comparisons across strongly varying base rates less straightforward than for the PSS. We therefore report HSS and CSI as complementary diagnostics rather than as primary measures of discrimination. In particular, their values are not directly comparable across substantially different base rates, since changes in event frequency can affect the attainable scores independently of the intrinsic discrimination ability of the model.
For both scores, the probability threshold used to obtain binary forecasts is selected numerically on the validation data using a dedicated automated procedure. 
%%---------------------------------------------------
%%CSI HSS and other wind
%%---------------------------------------------------
%\appendixfigures
\begin{figure*}[ht]
\centering
\includegraphics[width=0.7\linewidth]{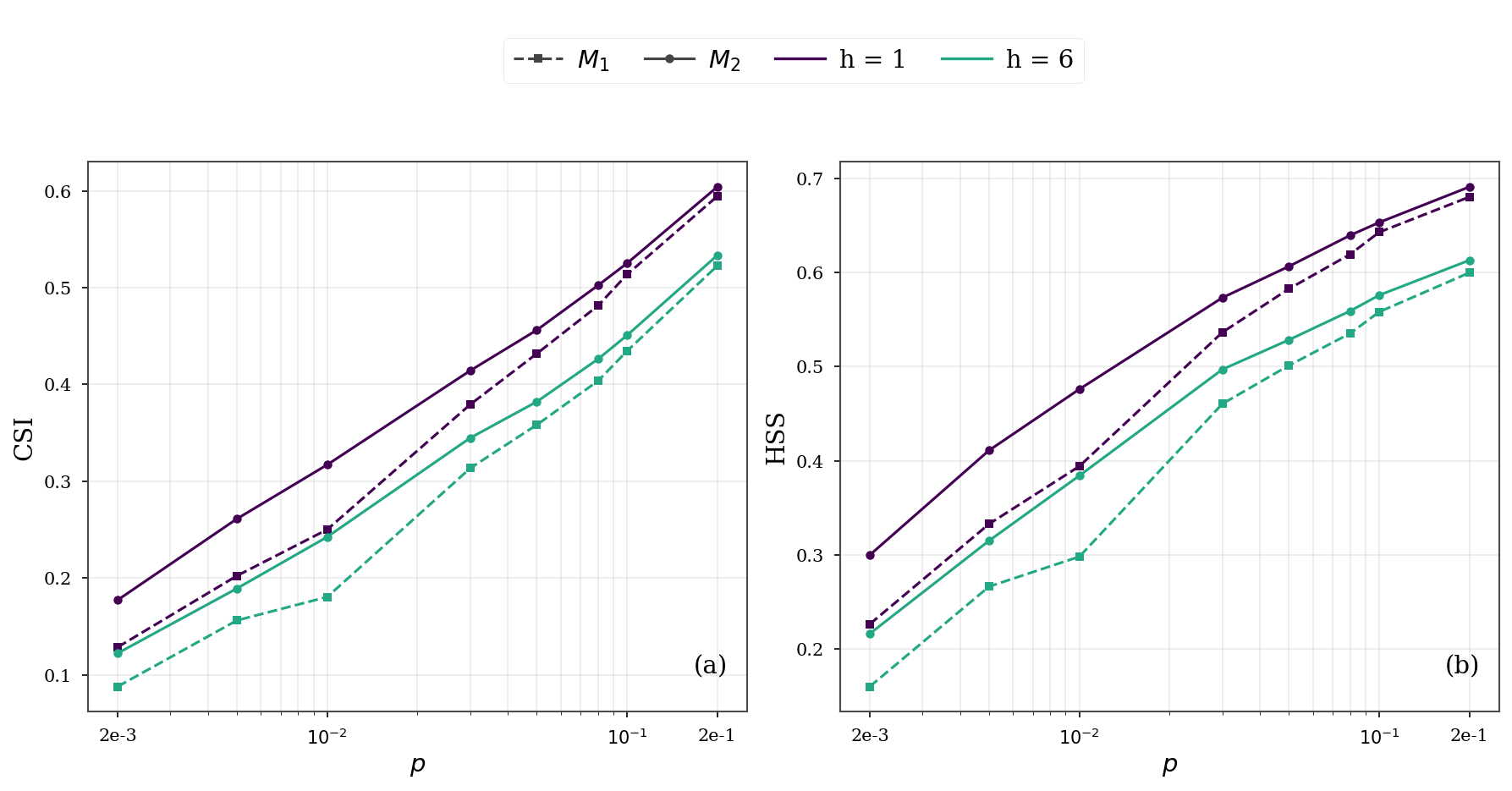}%{Figs/WIND_results_CSI_HSS.png}
\caption{ \footnotesize CSI (panel ~(a)) and HSS (panel ~(b))  relative to \textbf{ hourly wind speed} forecasts are displayed for models ${M}_1$ (symbols (\scalebox{0.7}{$\blacksquare$}) and dashed lines) and ${M}_2$ (symbols ($\bullet$) and solid lines). Two different forecast horizons are highlighted: $h=1$ h (in violet) and $h=6$ h (green).  }
\label{fig:Wind_CSI_HSS}
\vspace*{0cm}
\end{figure*}
%%---------------------------------------------------
%%CSI HSS and other rain
%%---------------------------------------------------
    \begin{figure*}[ht]
	\centering
	\includegraphics[width=0.7\linewidth]{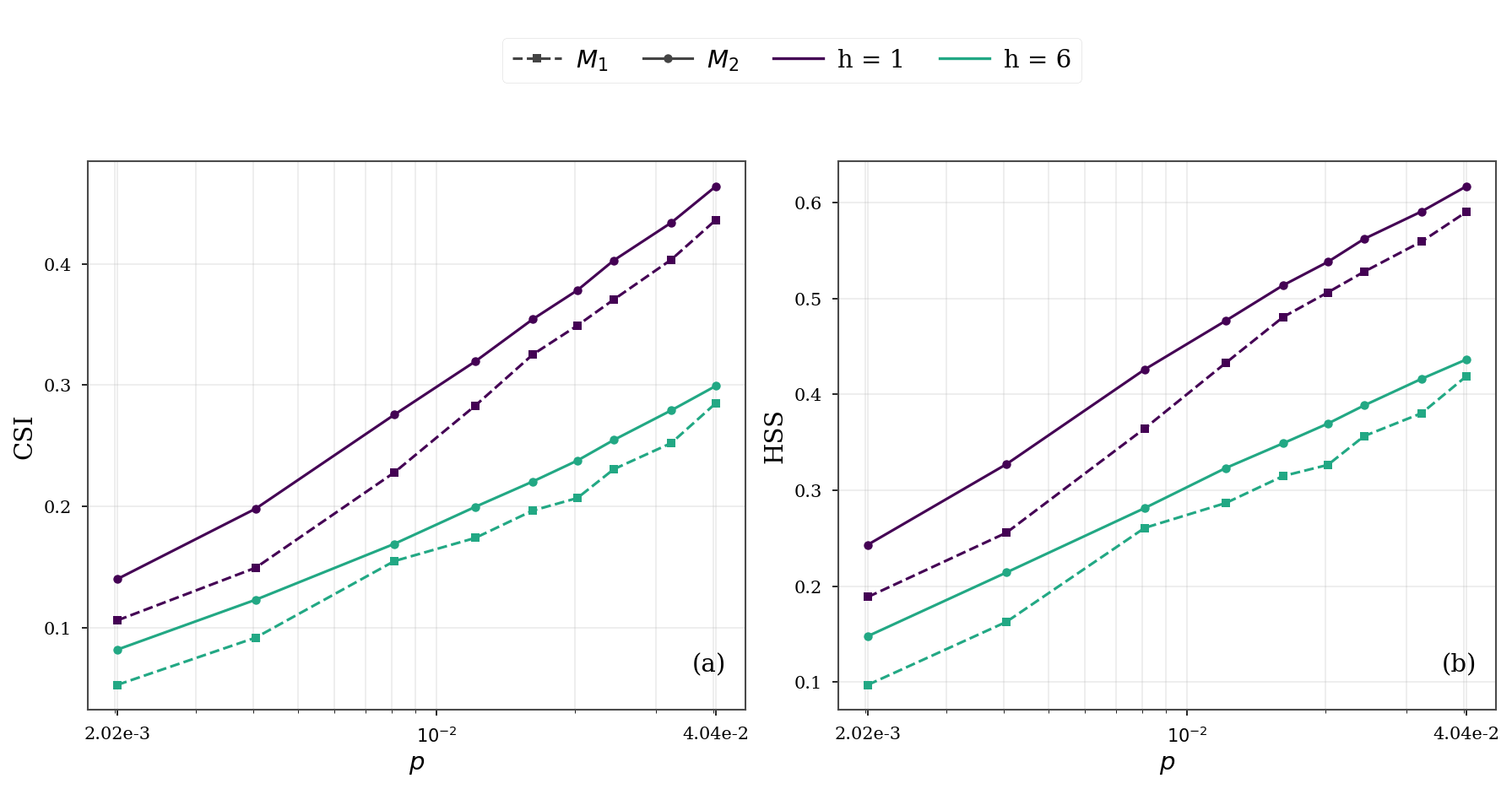}%{Figs/RAIN_results_CSI_HSS.png}
	\caption{\footnotesize CSI (panel (a)) and HSS (panel (b)) relative to \textbf{hourly accumulated rainfall} forecasts are displayed for the two models ${M}_1$ (symbols (\scalebox{0.7}{$\blacksquare$}) and dashed lines) and ${M}_2$ (symbols ($\bullet$) and solid lines). Two different forecast horizons are highlighted: $h=1$ h (in violet) and $h=6$ h (green). }
	\label{fig:RAIN_CSI_HSS}
    \vspace*{0cm}
\end{figure*}

Figures~\ref{fig:Wind_CSI_HSS} and~\ref{fig:RAIN_CSI_HSS} report CSI and HSS for wind speed and accumulated rainfall, respectively. For both variables, forecast horizons, and all considered exceedance probabilities, $M_2$ consistently  outperforms $M_1$. These complementary metrics therefore confirm the qualitative conclusion obtained from the primary verification measures, namely the superior performance of the distributional approach across increasingly rare thresholds.

\noappendix

\end{document}